\documentstyle[12pt,epsfig]{article}

\begin{document}

\def\be{\begin{equation}}
\def\ee{\end{equation}}
\def\br{\begin{eqnarray}}
\def\er{\end{eqnarray}}
\def\bc{\begin{center}}
\def\ec{\end{center}}
\def\ks {\not\!k}
\def\ps {\not\!p}
\def\qs {\not\!q}
\def\qqs {\not\!{\tilde q}}
\def\ds {\not\!\partial}
\def\eps {\not\!{\epsilon}_\lambda}
\def\dmu {\partial_{\mu}}
\def\dnnu {\partial^{\nu}}
\def\dmmu {\partial^{\mu}}
\def\dbeta {\partial_{\beta}}
\def\dbbeta {\partial^{\beta}}
\def\dlambda{\partial_{\lambda}}
\def\dllambda{\partial^{\lambda}}
\def\dnu{\partial_{\nu}}
\def\dnnu{\partial^{\nu}}
\def\d {\partial}
\def\xs {\not\!x}
\def\ps {\not\!p}
\def\ol{\overline}
\def\piNg{\pi N \gamma}
\def\piN{\pi N}
\def\M {{{\cal M}}}
\def\G {{{\cal T}}}
\def\T {{{\cal T}}}
\def\U {{{\cal U}}}
\def\V {{{\cal V}}}
\def\Lh{\hat{\cal L}}
\def\L {{{\cal L}}}
\def\ra{\rightarrow  }
\def\qv {\vec{q}}
\def\vv {\vec{v}}
\def\qvp{\vec{q}~'}
\def\qvpp {\vec{q}~"}
\def\pv {\vec{p}}
\def\pvp{\vec{p}~'}
\def\pvpp {\vec{p}~"}
\def\xv {\vec{x}}
\def\xvp {\vec{x}~'}
\def\tv {\vec{\tau}}
\def\elmu {\epsilon_{\lambda~\mu}}
\def\elaalpha {\epsilon_{\lambda}^{~\alpha}}
\def\elmmu {\epsilon_{\lambda}^{~\mu}}
\def\el {\epsilon_{\lambda}}
\def\kv {\vec{k}}
\def\NPi#1{{1\over\sqrt{2\omega(#1)}} }
\def\NN#1{\sqrt{{m_N \over \epsilon(#1)}} }
\def\Nk{{1\over \sqrt{2 \omega_{\gamma}(\kv)}}}
\def\N3{{1\over (2\pi)^3}}
\def\rf#1{{(\ref{#1})}}
\def\rfto#1#2{{(\ref{#1}-\ref{#2})}}
\def\E#1{E(#1)}
\def\W#1{\omega(#1)}
\def\wg{\omega_{\gamma}}
\def\ket#1{|#1 \rangle}
\def\bra#1{\langle #1|}
\def\ubar#1{\overline{u}(#1)}
\def\u#1{u(#1)}
\def\Top{\hat{T}}
\def\Mop{\hat{M}}
\def\Jop{\hat{J}}
\def\Mhop{\hat{{\hspace{-.8mm}\tilde M}}}
\def\Uop{\hat{U}}
\def\Vop{\hat{V}}
\def\tVop{\hat{\tilde V}}
\def\Mop{\hat{M}}
\def\Vhop{\hat{{\hspace{-.8mm}\tilde V}}}
\def\Gop{\hat{G}}
\def\Gammaop{\hat{\Gamma}}
\def\Gamopmu{\hat{\Gamma}_{\mu}}
\def\Gamopmunu{\hat{\Gamma}_{\mu\nu}}
\def\Gamopnumu{\hat{\Gamma}_{\nu\mu}}
\def\Gamopnumualpha{\hat{\Gamma}_{\nu\mu\alpha}}
\def\Gamopmunualpha{\hat{\Gamma}_{\mu\nu\alpha}}
\def\psix{\psi(x)}
\def\psixb{\bar{\psi}(x)}
\def\pix{\vec{\pi}(x)}
\def\pixx{\pi_3(x)}
\def\psidxmu{\psi_{\Delta \mu}(x)}
\def\psidxnu{\psi_{\Delta}_{\nu}(x)}
\def\psidxmmu{\psi_{\Delta}^{\mu}(x)}
\def\psidxnnu{\psi_{\Delta}^{\nu}(x)}
\def\psidxnu{\psi_{\Delta}_{\nu}(x)}
\def\psidxbmu{\bar{\psi}_{\Delta \mu}(x)}
\def\psidxbmmu{\bar{\psi}_{\Delta}^{\mu}(x)}
\def\psidxbnnu{\bar{\psi}_{\Delta}^{\nu}(x)}
\def\psidxbnu{\bar{\psi}_{\Delta \nu}(x)}
\def\psidxbalpha{{\bar{\psi}}_{ \Delta \alpha}(x)}
\def\psidxbaalpha{{\bar{\psi}}_{ \Delta}^{\alpha}(x)}
\def\rhoxmu{\vec{\rho}_{\mu}(x)}
\def\rhoxmmu{\vec{\rho}^{~\mu}(x)}
\def\rhoxnnu{\vec{\rho}^{~\nu}(x)}
\def\g5{\gamma_5}
\def\gmu{\gamma_{\mu}}
\def\gmmu{\gamma^{\mu}}
\def\gnu{\gamma_{\nu}}
\def\gnnu{\gamma^{\nu}}
\def\galpha{\gamma_{\alpha}}
\def\gaalpha{\gamma^{\alpha}}
\def\gmmunnu{g^{\mu\nu}}
\def\gnnummu{g^{\nu\mu}}
\def\gmualpha{g_{\mu\alpha}}
\def\gmmuaalpha{g^{\mu\alpha}}
\def\gnualpha{g_{\nu\alpha}}
\def\gnnuaalpha{g^{\nu\alpha}}
\def\smunu{\sigma_{\mu\nu}}
\def\gmunu{g_{\mu\nu}}
\def\gnumu{g_{\nu\mu}}
\def\snnubbeta{\sigma^{\nu\beta}}
\def\salphabeta{\sigma_{\alpha\beta}}
\def\fpimpi{\left({f_{\pi} \over m_{\pi}}\right)}
\def\fpiNdmpi{\left({f_{\piN\Delta} \over m_{\pi}}\right)}
\def\grhopipi{g_{\rho\pi\pi}}
\def\endauthors{}
\def\authors#1\endauthors{#1}
\def\pv {\vec{p}}
\def\ubar#1{\overline{u}(#1)}
\def\u#1{u(#1)}
\def\qv {\vec{q}}
\def\gnnu{\gamma^{\nu}}
\def\gmu{\gamma_{\mu}}
\def\gmmu{\gamma^{\mu}}
\def\gnu{\gamma_{\nu}}
\def\gnnu{\gamma^{\nu}}
\def\galpha{\gamma_{\alpha}}
\def\gaalpha{\gamma^{\alpha}}
\def\gmmunnu{g^{\mu\nu}}
\def\gnnummu{g^{\nu\mu}}
\def\gmualpha{g_{\mu\alpha}}
\def\gmmuaalpha{g^{\mu\alpha}}
\def\gnualpha{g_{\nu\alpha}}
\def\gnnuaalpha{g^{\nu\alpha}}
\def\smunu{\sigma_{\mu\nu}}
\def\gmunu{g_{\mu\nu}}
\def\gnumu{g_{\nu\mu}}
\def\snnubbeta{\sigma^{\nu\beta}}
\def\salphabeta{\sigma_{\alpha\beta}}
\def\Ps {\not\!P}
\def\ks {\not\!k}
\def\xs {\not\!x}
\def\gmmunnu{g^{\mu\nu}}
\vskip1.5cm

\bc
{\Large\bf  Elastic and radiative $\pi^+p$ scattering
and properties of the $\Delta^{++}$ resonance }
\ec
\vskip 2cm
\bc
{\large G. L\'opez Castro and A. Mariano} \\

\

{\it Departamento  F\'\i sica, Centro de Investigaci\'on y de
Estudios Avanzados} \\ {\it del IPN, Apdo. Postal 14-740, 07000 M\'exico,
D.F., M\'exico} \\

\ec
\vskip1.5cm
{ \large \bf Abstract}
\vskip 0.5 cm

We study the $\Delta^{++}$ contributions to elastic and radiative $\pi^+p$
scattering within an effective Lagrangian model which incorporates the
$\Delta,\ N, \ \rho$ and $\sigma$ meson degrees of freedom. This model
provides a description of the $\Delta$ resonance and its interactions
that respects electromagnetic gauge invariance and invariance under
contact transformations when finite width effects are incorporated.
Following recent developments in the description of unstable gauge
bosons, we use a complex mass scheme to introduce the finite width of
the $\Delta^{++}$ without spoiling gauge invariance. The total
cross section of elastic $\pi^+p$ scattering, whose amplitude
exhibits the resonant plus background structure of $S$-matrix theory, is
used to fix the mass, width and strong coupling of the $\Delta$ resonance.
The differential cross section of elastic scattering is
found in very good agreement with experimental data. The magnetic dipole
moment of the $\Delta^{++}$, $\mu_{\Delta}$,  is left as the only
adjustable parameter in radiative $\pi^+p$ scattering. From a fit to the
most sensitive configurations for photon emission in this process, we
obtain  $\mu_{\Delta} = (6.14 \pm 0.51)e/2m_p$, in agreement  with
predictions based on the SU(6) quark model.


\vskip1.5cm

\newpage

\begin{flushleft}
{ \large \bf 1. Introduction}
\end{flushleft}

\bigskip

  According to the most recent compilation of the Particle Data Group
\cite{pdg98},
the properties of the overwhelming majority of baryonic resonances are
badly known. Indeed, only conservative intervals are provided for most of
the masses
and widths of these resonances, either corresponding to the Breit-Wigner
formula (with energy-dependent width) or to the
pole position parameters \cite{pdg98}. Beyond masses and decay
widths, other parameters characterizing baryonic resonances are even less
known. In particular, a few determinations of the magnetic dipole moment
(MDM) are
provided only for the $\Delta^{++}$ resonance based on different
approaches to  radiative $\pi^+p$ scattering
\cite{Musa, Hell, wittman, Lin, pt}. Since the central values reported in
these references spread over a wide range, the Particle Data Group
\cite{pdg98} prefers to quote the interval $\mu_{\Delta} \sim 3.7$
to $7.5$ (in units of nuclear magnetons $e/2m_p$) for the $\Delta^{++}$
MDM. In the present paper, we are seeking for a determination of this
important property of the $\Delta^{++}$ in a full dynamical model which
consistently describes elastic and radiative $\pi^+ p$ scattering.

The different determinations of the $\Delta^{++}$ MDM reported in
\cite{pdg98} are based mainly on fits to the radiative $\pi^+p$
scattering data of Refs.  SIN \cite{sin} and UCLA \cite{ucla}
experiments (see also Refs. \cite{Arman, Leung}). Several
inconsistencies or limitations can be pointed out among the
different theoretical models used for this purpose. Some of these
difficulties can be traced back to the set of  Feynman rules used
to describe the propagator and the vertices involving the
$\Delta^{++}$ resonance. It is clear that any dynamical model
used to describe the $\Delta$ resonance must satisfy general
principles of quantum field theories such as gauge invariance and
invariance under contact transformations. Electromagnetic gauge
invariance is a fundamental property related to electric charge
conservation.
 On another hand, invariance under contact transformations ensures that
physical amplitudes involving the $\Delta$ resonance are independent of
any arbitrariness in the Feynman rules of a given theoretical
model for this resonance\cite{surdarshan,etemadi}. This invariance assures
that spurious spin-1/2 components are removed from the field describing an
on-shell $\Delta$ particle. However, the propagation of an off-shell
$\Delta$ particle unavoidable carries an spin-1/2 component as shown for
example in Ref. \cite{Benme}.
Thus, we must check that these properties are not
spoiled when the unstable character (finite width) of the
resonance is taken into account. As it has been pointed out
before, most of the current determinations of the $\Delta^{++}$
magnetic dipole moment \cite{pdg98} do not fulfill these general
requirements.

   In addition, the necessity of {\it ad hoc} form factors for vertices
involving the $\Delta^{++}$ are usually advocated to
achieve a better fit of the relevant experimental data.
According to the
philosophy of effective Lagrangian models, in the description of
low-energy hadron interactions  (as it is the case of elastic and
radiative $\pi^+p$ scattering in the $\Delta^{++}$ resonance region), we
must incorporate only the (structureless) relevant degrees of freedom at
that energy. Accordingly, we expect form factors to play an important role
only at higher energies. A typical example of such approach to the
radiative $\pi^+p$ scattering is provided by the non-relativistic isobar
model used in Refs.  \cite{Hell}. Thus, in the present paper we use an
{\it improved} Born approximation which is obtained by inserting the
propagator of an unstable $\Delta^{++}$ particle in the tree-level
amplitudes of elastic and radiative $\pi^+p$ scattering.


 Another kind of inconsistency has to do with the ambiguities related
to invariance under contact
transformations of Rarita-Schwinger fields $\psi^{\mu}(x)$. As is well
known, the Feynman rules involving the propagator and vertices of the
$\Delta^{++}$ depend upon an arbitrary parameter $A$ \cite{etemadi}. This
arbitrary parameter is the trace left in the Lagrangian of the model by
the contact transformations (see Eq. (4) below).
Decay amplitudes calculated from the $A$-dependent
Feynman rules should be, however, independent of $A$ \cite{etemadi}. A
common mistake in some calculations \cite{Schutz} is to take the simplest
form of the
propagator for the $\Delta^{++}$ corresponding to $A=-1$
and, simultaneously, the vertex $\pi N\Delta$ for a different value of
this parameter (for example $A=-1/3$). This inconsistency is also present
in some
determinations of the $\Delta^{++}$ MDM from $A$-dependent radiative
$\pi^+p$ scattering amplitudes; For example, the relativistic models of
Refs. \cite{Musa, wittman} use an specific
value of this arbitrary parameter. The determination of
the $\Delta^{++}$ MDM
provided in Ref. \cite{pt} is free of these ambiguities related to contact
transformations. Their method requires being able to detach the
decay process $\Delta^{++} \rightarrow \pi^+p\gamma$ out of the whole
radiative $\pi^+p$ scattering amplitude \cite{pt}. However, resonances
are  non-perturbative phenomena associated to the pole of the S-matrix
amplitude  \cite{smatrix} and one can not detach them from its production
or decay mechanisms.

A third approach currently used to determine the $\Delta^{++}$
MDM is
based on the soft
photon approximation \cite{low}. In the approach of Ref. \cite{Lin}, the
calculation of the radiative $\pi^+p$ scattering amplitude
up to terms of order zero in the photon energy $\omega_{\gamma}$ requires
a specific ansatz for the off-shell elastic amplitude. Though some terms
of order $\omega^0_{\gamma}$ coming from the proton MDM are included
\cite{Lin}, other terms in the amplitude corresponding to the
four-particle vertices $\pi^+p \Delta^{++} \gamma$
(see Figs. 1(e-f) in Ref. \cite{elamiri}) are
ignored. Furthermore, the finite width of the $\Delta^{++}$ which one
expects to be important in the resonance region, are also ignored in Ref.
\cite{Lin}. One of the goals of the present paper (see also Ref.
\cite{nos01}) is to go beyond those approximations.

An important and new element of the present paper concerns the issue of
gauge invariance of the amplitudes involving the $\Delta$ resonance
when its unstable character (finite width) is taken into account. A
consistent model of this charged resonance should provide amplitudes which
are gauge-invariant by themselves instead of {\it imposing} this property
to the total amplitude. The search for a consistent
description of resonances within quantum field theories has received much
attention recently, with the advent of precise measurements of the $Z^0$
gauge boson properties at LEP \cite{z0}. In particular, it has been point
out \cite{z0} that the naive introduction of the finite width effects in
the
Feynman rules involving the resonance can spoil gauge-invariance. Among
the different methods
proposed to cure problems associated to this gauge non-invariance in the
case of gauge-bosons, the {\it fermion-loop scheme} \cite{fls} provides
one of the simplest solutions. In the case of the $W$ gauge boson, this
approach implies that
using a propagator and  electromagnetic vertex of the $W$ boson which
includes the absorptive pieces of fermionic corrections allows to compute
amplitudes that are gauge-invariant when finite width effects are present.
Similar conclusions were drawn in Ref. \cite{bls} for the case of the
$\rho^{\pm}$ meson resonance.

   The expressions for propagators and electromagnetic vertices of charged
spin-1 resonances (for massless particles in loop corrections) can also
be obtained by replacing the bare mass $M_0^2$ by $M^2-iM\Gamma$ in the
lowest order expressions for these Feynman rules. This so-called {\it
complex mass scheme} offers a
simple prescription to obtain gauge-invariant amplitudes when finite width
effects are present. Based on the examples provided by the gauge boson
and $\rho$ meson resonances, in the present paper we will extend this
prescription to the vertices and the propagator of the $\Delta^{++}$
resonance. We expect this rule would follow from the explicit computation
of
the (finite) absorptive corrections to the corresponding lowest order
vertices of the $\Delta$ that involve $\pi N$ loops (see Refs.
\cite{fls,bls}). Our belief is based
on the powerful constraint imposed by electromagnetic gauge invariance
rather than on the specific dynamical model responsible for loop
corrections to Green functions of the $\Delta^{++}$.

  The aim of the present paper is to determine the MDM
of the $\Delta^{++}$ resonance by using a {\it full} dynamical model
which consistently describes the data on elastic and radiative $\pi^+p$
scattering. This model includes the $\Delta,\ N, \rho$ and $\sigma$
meson degrees of freedom (see for example \cite{Mariano}). By fixing the
free
parameters (mass, width and
coupling to $\pi^+p$ of the $\Delta$ resonance) of the model from
the total cross section of $\pi^+p$ scattering in
the resonance region, we are able to reproduce reasonably well the
experimental
data for the differential cross sections for elastic $\pi^+p$ scattering.
In addition, let us emphasize that the mass and width of the $\Delta^{++}$
required to fit the data on elastic $\pi^+p$ scattering, are consistent
with the $\Delta^{++}$ parameters defined from the pole position of the
elastic amplitude.
Our approach also provides an amplitude for radiative $\pi^+p$
scattering that is gauge-invariant under electromagnetism when the finite
width effects of the $\Delta^{++}$ resonance are taken into account. From
the experimental data about this process, we are able to determine the
MDM of the $\Delta^{++}$ from the photon configurations that are more
sensitive to this property.

Our paper is organized as follows. In sec. 2 we present the
dynamical  model which contains the degrees of freedom relevant for
$\pi^+p$ elastic scattering and discuss the procedure to introduce finite
width effects in a consistent way. In sec. 3 we introduce electromagnetic
interactions and compute the gauge-invariant $\Delta^{++}$
contribution to the amplitude of radiative
$\pi^+p$ scattering. In section 4
we determine the parameters of the $\Delta^{++}$ resonance from elastic
and radiative $\pi^+p$ scattering data. Finally our conclusions are
given in sec. 5.

\bigskip
\bigskip
\begin{flushleft}
{ \large \bf 2. Elastic pion-proton scattering}
\end{flushleft}
\bigskip

In this section we introduce the interaction Lagrangians relevant for the
description of $\pi N$ elastic scattering. We also compute the
$\Delta^{++}$ contribution to the  amplitude of elastic $\pi^+p$
scattering. A comparison with experimental data on the total and
differential cross sections is deferred to section 4.1.

As is well known, the total cross section of $\pi^+ p$
scattering in the
energy range $T_{lab} = 100 \sim 300$ MeV ($T_{lab}$ denotes the
kinetic energy of incident pions in the Lab system) exhibits a resonant
behavior around $T_{lab} \approx 175 $ MeV, corresponding to the
production of the $\Delta^{++}$,
$(I,J)=(3/2,3/2)$ resonance. The experimental cross section can
contain also   non-resonant background contributions which can
either, be parametrized in terms of a softly-varying function around the
resonance (see for example \cite{elamiri,bernicha}) or computed explicitly
in a given dynamical model. In this paper we adopt the second approach.

For definiteness, let us write the expression of the total cross
section for $\pi^+ p$ elastic scattering:

\br
\sigma_T &=& {(2\pi)^4 m_N^2 \over 4|\qv ~\E{\pv} - \pv~
\W{\qv}|}\int{d\qvp \over (2\pi)^3
\W{\qvp}}
\int{d\pvp \over (2\pi)^3\E{\pvp}}  \delta^4(p+q -p'- q')
\nonumber \\
& & \times {1\over2}
 \sum_{ms',ms}  \left|
\M[\pi^+ p \rightarrow \pi^+ p](q',p',ms';q,p,ms)
\right|^2,\label{0}
\er
where $q= (\omega, \qv )$, $p = (E, \pv)$ denote pion and proton
four-momenta (primes refer to the same quantities in the final state),
respectively, $ms$ is the proton's spin projection, and
$\M[\pi^+ p \rightarrow \pi^+ p]$ denotes the Lorentz-invariant
scattering amplitude.

The dynamical model we advocate in this paper to compute the
scattering amplitude involves nucleons, the $\Delta^{++,\ 0}$ resonances,
the $\rho$ and the $\sigma$ meson degrees of freedom\footnote{Throughout
this
paper we will assume isospin symmetry in the masses and strong couplings
of hadrons.}. The various pieces of the interaction Lagrangian density
contributing to $\pi^+p$ scattering are (see for example \cite{Mariano})
\footnote{The Lagrangian densities
corresponding to kinetic terms can be found for example in Ref.
\cite{elamiri}.} :

\br
\Lh_{hadr} = \Lh_{\pi NN} + \Lh_{\Delta \pi N} + \Lh_{\rho NN} + \Lh_{\rho
\pi\pi} + \Lh_{\sigma NN} + \Lh_{\sigma \pi\pi}
\nonumber
\er
where the individual terms are given by
\br
\Lh_{\pi NN}(x) & = & -\left({f_{\pi N N} \over m_{\pi}}\right) \psixb \g5
\tv.(\ds\pix) \psix ,
\nonumber
\\
\Lh_{\Delta \pi N}(x) & = & \left({f_{\Delta\piN} \over
m_{\pi}}\right)
\psidxbmmu \Lambda_{\mu\nu}(A) \vec{T}^{~\dag}.(\dnnu \pix)\psix
+ h.c.
,
\nonumber
\\
\Lh_{\rho NN}(x)& = &-{1 \over 2}g_{\rho}\psixb  \left[ \gmu -
{\kappa_{\rho} \over 2 m_N}
\smunu \dnnu\right]\tv.\rhoxmmu\psix, \nonumber
\\
\Lh_{\rho \pi\pi}(x)& = & - \grhopipi  \rhoxmu.(\pix \wedge \dmmu
\pix),\nonumber
\\
\Lh_{\sigma NN }(x)& = & g_{\sigma NN}\psixb \psix \sigma(x),\nonumber
\\
\Lh_{\sigma \pi\pi}(x)& = & \left({g_{\sigma \pi \pi} \over 2
m_\pi}\right)\sigma(x)(\dmu \pix) .(\dmmu \pix), \label{2}
\er
and ($A$ is an arbitrary parameter related to contact transformations to
be defined below),
\br
\Lambda_{\mu\nu}(A) = g_{\mu\nu} + {1 \over 2} (1 + 3 A)\gamma_\mu
\gamma_\nu \ .\nonumber
\er
The isotopic fields $\psix $ and $\psidxmmu$ denote the $N$ and $\Delta$
baryons, respectively, while $\pix$, $\rhoxmmu$ and $\sigma(x)$ denote
the pion, $\rho$-  and $\sigma$-meson fields.
The arrow over the meson fields refers to the isospin space.
$\vec{T}^{~\dag}$ stands for the isospin $1/2$ to $3/2$ transition
isospin operators, while $\vec{\tau}$ is used for Pauli isospin matrices.
The constants  $f_{\pi N N}$, $g_{\sigma
NN}$,  $f_{ \Delta\pi N}$, $\grhopipi$,  and $g_{\sigma \pi \pi}$ denote
the strong couplings among the different particles indicated as
subindices, and $g_{\rho}=g_{\rho \pi\pi}$ ($\kappa_{\rho}$) is the vector
(magnetic) coupling of the $\rho$-meson to the nucleon.

Note  that $\Lh_{\Delta \pi N}(x)$ in Eq.\rf{2} is invariant
under the contact transformation\footnote{The kinetic term for the
spin-3/2 field and other interaction Lagrangians involving the
$\psi^{\mu}(x)$ in Eq. (9) also remain invariant under this
transformation (see ref. \cite{etemadi, elamiri}).}:
\br && \psi^{\mu}
\rightarrow \psi^{\mu} + a\gamma^{\mu}\gamma_{\alpha}\psi^{\alpha}\ ,
\ \ \ \ A \rightarrow A' = {A -2a\over 1+4a},\label{3}
\er
where $a$ is an arbitrary parameter (excluding the value $a = -1/4$).
Physical scattering amplitudes
must be independent of any particular choice of $A$ in vertices and
propagators involving the $\Delta$ baryon. As already
advertised in the introduction, this fact is sometimes ignored in
computing the $\Delta^{++}$ contribution to elastic and radiative
$\pi^+p$ scattering amplitudes (see for example Refs.\cite{Musa,
wittman}).

  In the present model, the elastic scattering amplitude is composed of
several pieces:
\be
{\cal M} (\pi^+p \rightarrow \pi^+p) = \sum_{i=\Delta^{++}, \Delta^0, n,
\rho,\sigma} {\cal M}_{i}(\pi^+p \rightarrow \pi^+p)\ .\label{4b}
\ee                                                            
The contributions to the  amplitude involving intermediate
nucleon and $\Delta^0$ baryons (in the crossed-channel) and the $\rho$ and
$\sigma$ mesons (in the $t$-channel) are shown in Figure 1. They will
be included at the {\it tree-level} since they provide a smoothly-varying
background term in the amplitude
around the resonance region. The evaluation of these amplitudes  from the
interaction Lagrangians given above is straightforward.
Therefore, we will focus now in the explicit form
of the $\Delta^{++}$ contribution (resonance in the $s$-channel).

As it was proven in Ref. \cite{elamiri}, in the case of elastic and
radiative $\pi^+ p$ scattering, the $A$-dependent Feynman rules
involving the $\Delta$ can be replaced by a set of $A$-independent
vertices and propagators called {\it reduced} Feynman rules.
The calculation of $\pi^+p$
scattering amplitudes using either set of rules, give rise to identical
results \cite{elamiri}. The elastic scattering with the $\Delta^{++}$
intermediate
state in the $s$-channel (see the first graph in Figure 1) is given by
\begin{equation}
{\cal M}_{\Delta^{++}}(\pi^+ p \rightarrow \pi^+
p) =
-i \left({f_{\Delta
\pi
N}
\over
m_{\pi}}\right)^2
\ubar{p'} q'_\mu G^{\mu\nu}(p+q) q_\nu \u{p},\label{4}
\end{equation}
where $u(p) \equiv u(p,ms)$ denote proton spinors, and the reduced form of
the  $\Delta^{++}$ propagator is given by
\begin{eqnarray}
G^{\mu\nu}(P)  &=&  {i \over P^2 -m_\Delta^2}
\left\{ (\not P + m_{\Delta}) \left [-\gmmunnu + \frac{1}{3} \gmmu \gnnu +
\frac{2}{3} {P^\mu P^\nu \over
m_\Delta ^2 }- \frac{1}{3} {P^\mu \gnnu - P^\nu \gmmu \over
m_\Delta }\right] \right. \nonumber \\
& & \left. - {
2\over 3 m_\Delta ^2}(P^2 - m_\Delta ^2)\left[{ \over
}(\gmmu
P^\nu - \gnnu  P^\mu) + (\Ps + m_\Delta)\gmmu \gnnu\right] \right\}.
\label{5}
\end{eqnarray}

  The elastic  amplitude of Eq. (5) blows up when the total energy
$\sqrt{P^2}$ approaches
$m_{\Delta}$. As is well known \cite{pdg98}, this bad behavior can be
cured naively by replacing $m_{\Delta}^2 \rightarrow
m_{\Delta}^2-im_{\Delta}\Gamma_{\Delta}$ in the
denominator of the $\Delta^{++}$ propagator\footnote{The finite widths of
the $\Delta^0$ baryon and the $\rho,\ \sigma$ mesons do
not play any role since these resonances do not appear
in the $s$-channel.}($\Gamma_{\Delta}$ is
the decay width of the $\Delta$). This procedure leads to a
modified (or {\it improved}) Born amplitude. However, as it has been
widely discussed
elsewhere \cite{fls}, this prescription breaks gauge invariance when
photons are
attached to all charged particles of the elastic scattering
(see the first row in Figure 2) to produce the corresponding
radiative decay amplitude.

   The simplest solution to maintain gauge invariance in the radiative
$\pi^+p$ scattering amplitude when finite width effects are incorporated,
is to replace $m_{\Delta}^2  \rightarrow
m_{\Delta}^2-im_{\Delta}\Gamma_{\Delta}$ in {\it all} the Feynman
rules involving the $\Delta^{++}$ resonance \cite{elamiri} (this is the
so-called  complex mass scheme). As it has been explained in the
introduction, this prescription is well justified for the amplitudes
involving intermediate $Z^0$ and $W^{\pm}$ gauge boson \cite{w,pn93} and
the $\rho^{\pm}$ meson resonances \cite{bls}. This is a rigorous result
obtained when
one-loop absorptive corrections are included in the propagator and the
electromagnetic vertex of the charged resonance, with massless particles
running
in loops \cite{fls,bls}. Thus, in the present case we will use the
complex mass scheme as a useful {\it prescription} to obtain
gauge-invariant
amplitudes. As in the case of spin-1 charged resonances, we expect that
absorptive pieces of $\pi^+p$ loop corrections to the $\Delta^{++}
\rightarrow \gamma   \Delta^{++}$ vertex and the $\Delta^{++}$ propagator
would generate
these results in the limit of massless pions and protons. Since this limit
may me questionable, let us point out that a Laurent expansion of
the amplitude (with Dyson summation of the $\Delta^{++}$ self-energy
corrections ) around the pole  will give similar
results to the complex mass scheme, with regular terms being absorbed into
background contributions (see for example Refs. \cite{elamiri,z0,w} and
the discussion we present in the appendix).


 Finally, let us mention that within our modified Born approximation the
non-resonant (background) amplitudes are real since they are included at
the tree-level. This will produce a lack of unitarity of the
total amplitude for the elastic scattering in our model. In the Appendix
we have included a discussion about how the contributions we have
neglected in our approximation would help to restore unitarity. In
particular, we have found that the value extracted for the magnetic dipole
moment of the $\Delta^{++}$ is not sensibly affected by our
approximations.

\bigskip
\bigskip
\bigskip

\begin{flushleft}
{\large\bf 3. Radiative pion-proton scattering }
\end{flushleft}

\bigskip

In this section we focus on the calculation of the gauge-invariant $\pi^+p
\rightarrow \pi^+ p \gamma$ scattering amplitude in the presence of
the $\Delta^{++}$ finite width effects. Comparison with experimental data
on energy and angular photon distributions is left to section 4.2.

For definiteness, let us start with the expression for the differential
cross section of this process:
\br
{d\sigma \over d\omega_{\gamma}d\Omega_{\pi}d\Omega_{\gamma}}
&=&
{(2\pi)^4\wg m_N^2 \over 8|\qv ~\E{\pv} - \pv~
\W{\qv}|}
\int{d|\qvp||\qvp|^2 \over (2\pi)^3\W{\qvp}}
\int{d\pvp \over (2\pi)^3\E{\pvp}}  \delta^4(p + q - p' - q' - k)
\nonumber \\
& & \times {1\over2}
 \sum_{\el,ms',ms}  \left|
\M[\pi^+p \rightarrow \pi^+ p
\gamma](\el,k;q',p',ms';q,p,ms)
\right|^2,\label{6}
\er
where $k =(\omega_{\gamma}, \kv)$ and $\el$ are the  four-momentum and
 polarization four-vector of the photon, respectively. In section
4.2, we will compare this cross section calculated in our model with a set
of experimental data\footnote{ In particular, this observable will be
represented as a function of the photon energy $\omega_{\gamma}$ for fixed
angular configurations of photons and pions emitted in the process.}
 in order to extract a value for the magnetic dipole moment of the
$\Delta^{++}$ baryon resonance.

  The amplitude for radiative $\pi^+p$ scattering can be
obtained by attaching the photon to all external and internal lines in
Feynman diagrams of elastic scattering (shown in Figure 1). In addition,
we will also have
four-point vertices involving the photon (see Figure 2). Such vertices
arise from the derivative couplings of baryons and mesons in Eqs.
(3). The Feynman rules required to
introduce electromagnetism are derived from the following effective
Lagrangian:

\be
\Lh_{elec} = \Lh_{\gamma NN} + \Lh_{\gamma \pi NN} +
\Lh_{\gamma\pi\pi}
   + \Lh_{\gamma \Delta\Delta} + \Lh_{\gamma \pi N \Delta}
+ \Lh_{\gamma\rho\pi\pi} + \Lh_{\gamma\sigma\pi\pi} \ .
\label{7}
\ee
The different pieces of this Lagrangian are the following:
\br
\Lh_{\gamma NN}(x) & = & - e \psixb \left[\gmu -
{\kappa_p \over 2 m_N} \smunu \dnnu \right ]  A^{\mu}(x) \psix
,\nonumber \\
\Lh_{\gamma\pi NN}(x) & = & - e \left({f_{\pi N N} \over
m_{\pi}}\right)
\psixb \g5 \gmu \left[ \tv \times \pix \right ] _3 \psix A^{\mu}(x),
\nonumber \\
\Lh_{\gamma \pi\pi}(x) & = &  - e\left[\pix \times \dmu \pix
\right]_3
A^{\mu}(x), \nonumber \\
\Lh_{\gamma \Delta\Delta}(x) & = & - 2e \psidxbnu
\Lambda^{\nu\nu'}(A)\Gamma_{\nu'\mu'\alpha}\Lambda^{\mu'\mu}(A)
\psidxmu
A^{\alpha}(x)
,\nonumber \\
\Lh_{\gamma \Delta \pi N}(x) & = &
e\left({f_{\Delta\pi N} \over m_{\pi}}\right)
\psidxbmmu \Lambda_{\mu\nu}(A) \left [\pix \times
\vec{T}^{~\dag}\right]_3 \psix
A^{\nu}(x)+ h.c.,
 \nonumber
\\
\Lh_{\gamma\rho\pi\pi}(x) & = & e g_\rho \left\{[\pix.\pix - \pixx
\pixx]
\rho^\nu_3(x)\right.\nonumber \\
& &\left. - \pixx[\pix.\rhoxnnu - \pixx
\rho^\nu_3(x)]\right\}A_{\nu}(x),
\nonumber
\\
\Lh_{\gamma \sigma \pi\pi}(x) & = &  - 2e
\left({g_{\sigma \pi \pi} \over 2 m_\pi}\right)\sigma(x) \left[
\pix \times \dmu \pix\right]_3
A^{\mu}(x),
\label{8}
\er
where $A_\nu(x)$ is  the electromagnetic four-potential and
$e$ ($\kappa_p$) denotes the charge (anomalous magnetic moment) of the
proton. The tensor $\Lambda_{\mu\nu}(A)$ has been defined in the previous
section. The interaction Lagrangians of Eq. (9) involving the $\Delta$ are
invariant under the contact transformations of Eq. (4).

The electromagnetic vertex of the $\Delta^{++}$ resonance is given
by \cite{etemadi,elamiri}:
\begin{eqnarray}
\Gamma_{\alpha\beta\rho} & = &
\left( \gamma_{\rho}-\frac{i\kappa_{\Delta}}{2m_{\Delta}} \sigma_{\rho
\sigma}k^{\sigma} \right) g_{\alpha \beta} -\frac{1}{3} \gamma_{\rho}
\gamma_{\alpha} \gamma_{\beta} -\frac{1}{3} \gamma_{\alpha}g_{\beta \rho}
+ \frac{1}{3} \gamma_{\beta} g_{\alpha \rho}\ ,
 \label{10}
\end{eqnarray}
where we have allowed for a term describing the anomalous magnetic moment
$\kappa_{\Delta}$ of the $\Delta$ resonance. This quantity is related to
the total magnetic dipole moment of the $\Delta^{++}$ through the
following equation:
\begin{equation}
\mu_{\Delta}=2(1+\kappa_{\Delta}) \frac{e}{2m_{\Delta}} .
\end{equation}
Is is straightforward to verify that the vertex of Eq. (10) satisfy the
following Ward identity:
\begin{equation}
 G^{\mu\alpha}(P') \Gamma_{\alpha \beta \rho}  k^\rho
G^{\beta\nu}(P) = G^{\mu\nu}(P)-G^{\mu\nu}(P')
\end{equation}
where $P=P'+k$.
This identity ( together with the inclusion of {\it all} possible
radiative
diagrams)
guarantees that the radiative $\pi^+p$ scattering amplitude computed with
the Feynman rules of Eqs. (3) and (9) is automatically
gauge-invariant, even when the complex mass scheme is implemented.

   The interaction Lagrangians shown in Eqs. (3) and (9) give rise to the
Feynman diagrams displayed in Figure 2. The first line of Fig. 2
contains the contributions with the $\Delta^{++}$ as intermediate state.
The other two lines in Figure 2 correspond to the contributions with
intermediate
$\Delta^0,\ N$ and $\rho- \sigma$ mesons. The evaluation of these
gauge-invariant contributions is straightforward. Therefore, as  in the
case of elastic scattering, we will focus on the amplitude corresponding
to the $\Delta^{++}$ exchange in the $s$-channel. This amplitude
receives contributions from the  seven
Feynman graphs appearing in the first
row of Figure 2; it is given by \cite{elamiri}:
\begin{eqnarray}
\hspace{-1cm}{\cal M}_{\Delta^{++}}(\pi^+p \rightarrow \pi^+p \gamma) & =
&
 -i e \left({f_{\Delta \pi N}\over m_{\pi}}\right)^2 q'_\mu q_\nu
\ubar{p'}
\left[G^{\mu\nu}(P')\left({ q \cdot \epsilon
\over q \cdot k} +
{p\cdot \epsilon - R(p)\cdot \epsilon \over p \cdot k}\right)
\right. \nonumber \\
&-& \left. \left({ q' \cdot \epsilon
\over q' \cdot k} +
{p'\cdot \epsilon - R(p')\cdot \epsilon \over p' \cdot k}\right)
G^{\mu\nu}(P) + 2  G^{\mu\alpha}(P') \Gamma_{\alpha \beta \rho}
\epsilon^\rho G^{\beta\nu}(P)\right.\nonumber \\
& + & \left.{1 \over q \cdot k}G^{\mu\rho}(P')(\epsilon_\rho k^\nu -
\epsilon^\nu k_\rho) -
{1 \over q' \cdot k}(\epsilon^\mu k_\rho -
\epsilon_\rho k^\mu)G^{\rho\nu}(P) \right ]u(p)
\label{9}
\end{eqnarray}
where
\begin{eqnarray}
& &\hspace{-3.5cm} R_\mu(x) \equiv {1\over 4} [\ks,\gamma_\mu] +
{\kappa_p\over 8 m_N}
\{ [\ks,\gamma_\mu],\xs\},\nonumber
\end{eqnarray}
and $P = p+q $, $
P'=p'+q'$, such that $P=P'+k$. This amplitude exhibits the correct low
photon energy behavior expected by Low's theorem \cite{low}.

 As required, this amplitude  is explicitly gauge-invariant ({\it i.e.},
it vanishes when $\epsilon \rightarrow k$ by virtue of the Ward identity
given in Eq. (12)) and does not depend on the
parameter $A$ associated to contact transformations. As it was
discussed in the previous section, finite width effects of the
$\Delta^{++}$ are necessary in order to avoid that the amplitude
diverges when $P^2$ or $P'^2$ approaches $m_{\Delta}^2$. According to the
prescription of the complex mass scheme, we can perform the replacement
$m_{\Delta}^2 \rightarrow m_{\Delta}^2-im_{\Delta}\Gamma_{\Delta}$ in the
propagator and the electromagnetic vertex of the $\Delta^{++}$ without
destroying gauge-invariance.

\begin{flushleft}
{ \large \bf 4. Comparison with experimental data}
\end{flushleft}

 In this section we compare our dynamical model with experimental data on
elastic and radiative $\pi^+p$ scattering. We are not interested in
performing a global fit to the whole universe of experimental data about
these
processes. Our first goal in section 4.1 is to test how our model works in
the case of elastic $\pi^+p$ scattering. By using a reliable set of
experimental data on the total cross section for this process, we fix the
free parameters of our
model: the mass and width of the $\Delta$ and its coupling to $\pi^+p$.
With this information at hand, we will compare the {\it
prediction} of our model with
available experimental data on the differential cross section of
elastic $\pi^+p$ scattering. Finally, in section 4.2 we use a set of
experimental data on radiative $\pi^+ p$ scattering to extract the value
of the magnetic dipole moment of the $\Delta^{++}$, which is left as the
only free parameter of the model in this case.

\bigskip
\begin{flushleft}
{\large \it 4.1 Elastic $\pi^+p$ scattering: fixing the mass and  width of
the
$\Delta$}
\end{flushleft}

\bigskip

  The different contributions to the elastic $\pi^+p$ scattering amplitude
are shown in Fig.1. The amplitude corresponding  to the $\Delta^{++}$
resonance in the $s$-channel was given explicitly in Eqs. (5) and (6).
This is the most
important contribution to elastic scattering in the resonance region
($T_{lab} \approx 175$ MeV).
All other contributions should be added
coherently,
and the total amplitude $\M[\pi^+ p \rightarrow \pi^+ p]$ exhibits the
resonant plus background structure
expected by general principles of $S$-matrix theory \cite{smatrix}.
Finally we use Eq.\rf{0} to evaluate the total elastic cross section.

Some of the couplings that enter the interaction Lagrangians given in
Eq.\rf{2} are taken from other low energy processes. For example,
the coupling constants $(g_{\rho}^2, g_{\pi NN}^2)/4\pi = (2.9, 14)$
were taken from the $\rho \pi\pi$ decays and the analysis of NN
scattering data \cite{Schutz,sigma}, whereas the magnetic $\rho NN$
coupling
$\kappa_{\rho}=3.7$ was extracted from values of nucleon magnetic moments.
The masses of the $\rho$ meson and the nucleon were taken from
\cite{pdg98}, and the
mass of the hypothetical $\sigma$ meson was set to 650 MeV
\cite{Schutz,sigma}
(see the end of this section for other choices of this mass).
Thus, the mass and width of the $\Delta$ and the coupling
constants $g_{\sigma \pi\pi , \sigma NN}$ and $f_{\Delta N\pi}$
are left as the only free parameters to be determined from the
total cross section of $\pi^+p$ scattering.

In order to compare the size of the different contributions, we have
fitted the total cross section by adding successively, a new background
contribution (diagrams with $\Delta^0,\ N, \ \rho$ and $\sigma$ in
intermediate states) in each fit.  The results of these fits are shown in
Table 1. The curves corresponding to every one of the best fits to the
experimental cross sections of Ref.  \cite{pedroni} are plotted in Figure
3, for energies of incident pions in the range $75\ {\rm MeV} \leq T_{lab}
\leq 300\ {\rm MeV}$.

 Once we have fitted the free parameters relevant for elastic $\pi^+ p$
scattering, we can give an additional non-trivial check of the model. This
corresponds to the prediction of the differential cross section
$d\sigma/d\Omega_{\pi}$ as a function of the scattering pion angles
$\theta$, for fixed energies of incident pions. In Figure 4, we have
compared our predictions for the differential cross section with some
available experimental data \cite{Bussey, Sadler, Gordeev} for $T_{lab}
\approx 263$ and $291$ MeV, in a wide interval of scattering angles.

  Some interesting features are worth to be stressed from the fits to the
total and differential cross sections of elastic $\pi^+p$ scattering.
 First, we can observe that agreement with data in both cases is improved
when the contributions from all intermediate states ($\Delta^{++},\
\Delta^0,\ N,\ \rho$ and $\sigma$) are included. The values of
$\chi^2/dof$ are reasonably good given the uncertainties
inherent to the model\footnote{ The
$\chi^2/d.o.f.$ obtained in the last fit involving all the contributions,
drops to 4.5 when the last three points in the upper tail of the
total cross section are excluded.}(mainly related to the uncertainty in
the mass $\sigma$ meson meson).
In particular, the
contribution from the $\sigma$ meson allows to achieve a better fit to the
total cross section for $T_{lab}$ below the resonance peak (see Fig. 3).
It also allows to improve the agreement with the differential cross
section (see Figure 4) in the whole angular interval considered here.
Second, the values obtained for the mass and width of the $\Delta$'s,
namely $m_{\Delta}=(1211.2 \pm 0.4)$ MeV and $\Gamma_{\Delta}=(88.2 \pm
0.4)$ MeV,  are similar to those obtained from a model-independent
analysis of the same data on the total cross section, namely
\cite{bernicha}: $M=(1212.20 \pm 0.23)$ MeV and $\Gamma=(97.06 \pm 0.35)$
MeV\footnote{As is well known, the parameters defined from the pole
position are smaller than the ones obtained using a Breit-Wigner formula
with energy-dependent width
\cite{pdg98}.}. As it is discussed in the Appendix, this difference
probably arise from the non-unitarity implicit in our model. This
comparison provides a consistency check for our
model as long as both results corresponds to the mass and width parameters
defined from the pole position of the scattering amplitude. Note, however,
that this agreement is non trivial because in our model the background
contributions are {\it fixed} from independent low-energy phenomenology.
This is not the case within the $S$-matrix approach to the $\pi^+p$
amplitude
\cite{bernicha}, where the background terms are taken as slowly varying
functions to be fitted to experimental data. In other words, contributions
to the elastic amplitude with $\Delta^0,\ N, \ \rho$ and $\sigma$, indeed
simulate well the background terms obtained from the model-independent
analysis of Ref. \cite{bernicha}.

It is interesting to observe in Figure 4 that the model describes well the
differential cross section  despite the fact that these data correspond to
kinetic energies of incident pions ($T_{lab} \approx 263$ and $291$
MeV) that lie at the upper tail of the $\Delta^{++}$ resonance shape (see
Figure 3). This is very important because
incident pions in  radiative $\pi^+p$ scattering, to be considered below,
correspond to those particular values of kinetic energies.

 Finally one may wonder how the particular value of the $\sigma$ meson
mass ($m_\sigma = 650$ MeV) could have influenced the fitting procedure.
We have repeated the fits to the total cross section data by allowing
$m_{\sigma}$ to vary 200 MeV around its central value. The results are
shown in Table 2 and Figure 5 for the total cross section. Figure 6 shows
the corresponding predictions for the differential cross section. From
Table 2 and Figure 5, we observe that the mass of the $\sigma$ meson is
mainly correlated with its coupling constants to pions and nucleons, while
the changes in the $\Delta$ parameters are negligible. On the other
hand, the effects of these  variations in the mass of the $\sigma$
meson can be more important in the differential cross section of
elastic $\pi^+p$ scattering. Note however that our prediction for this
observable becomes worse for $\cos \theta >0$ when $m_{\sigma}$ deviates
from 650 MeV (see Figure 6), although it may slightly improve the
prediction for negative values of $\cos \theta$. Consequently, in the
following we will take $m_{\sigma}= 650$ MeV in our fitting procedure.

\begin{flushleft}
{\large \it 4.2 Radiative $\pi^+p$ scattering: determination of
$\mu_{\Delta}$}
\end{flushleft}

\bigskip

  The contributions to the radiative $\pi^+p$ scattering amplitude are
shown in Figure 2. The scattering amplitudes corresponding to the last two
rows in Fig. 2 can be computed in a straightforward way. The amplitudes
involving the $\Delta^{++}$ resonance, first row in Figure 2, have to take
into account the finite width effect of this resonance, preserve gauge
invariance and remain invariant under contact transformations. This
amplitude has been given explicitly in Eq. (13).

  Based on the analysis presented in the previous section, we can
easily check that the only free parameter in our description of radiative
$\pi^+p$ scattering is the $\Delta^{++}$ magnetic dipole moment. Thus, a
fit to the corresponding radiative data   should provide us with a
determination of this important property.

  In order to justify our choice of experimental data, let us remember
the structure of photon radiation in $\pi^+p$ scattering. As is well
known \cite{low}, the low photon-energy structure of the unpolarized
probability of radiative processes is given by:
\begin{equation}
\sum_{\epsilon_\lambda}|{\cal M}(\pi^+p \rightarrow \pi^+p\gamma)|^2 =
\frac{A}{\omega_{\gamma}^2} + B\omega_{\gamma}^0 + \cdots
\end{equation}
The terms of $O(\omega_{\gamma}^{-2})$ arise from photons radiated off
external charged particles and $A$ depends only on the parameters
of elastic $\pi^+p$ scattering. Terms of order $\omega_{\gamma}^{-1}$ are
absent in the
unpolarized probability by virtue of the Burnett and Kroll's theorem
\cite{bk}. The effects of the $\Delta^{++}$ magnetic dipole moment enter
this probability at order $\omega_{\gamma}^0$. Thus, for low-energy
photons, we expect the unpolarized probability to be sensitive to the
effects of the $\Delta^{++}$ MDM when some kinematical mechanisms allows
to suppress the leading dominant term (of $O(\omega_{\gamma}^{-2})$).
Fortunately, this situation occurs when photons are emitted in a backward
configuration to other charged particles \cite{kpz}. Actually, this has
been the motivation to study those particular geometrical configurations
 of radiative $\pi^+p$ scattering in Ref. \cite{ucla}.

Thus, the experimental data of Ref. \cite{ucla} are suitable for a
determination of the $\Delta^{++}$ magnetic dipole moment within our
model. Furthermore,
since we expect our approximation to be more reliable for low-energy
photons (where the electromagnetic static properties play the main role),
we will chose a subset of data corresponding to 20 MeV$ \leq
\omega_{\gamma} \leq$ 100 MeV. Following Ref. \cite{ucla} we label with
G1, G4
and G7 the geometrical configurations corresponding to photons detected at
angles $(\theta_{\gamma}, \phi_{\gamma})=(160^o,0^o),\ (140^o,0^o)$ and
$(120^o,0^o)$, respectively; {\it i.e.} the photons emitted backwards to
the outgoing pion, which in turn is emitted at angle $(\theta_{\pi},
\phi_{\pi})=(50.5^o,180^o)$ with respect to the direction of incident
pions. As already explained, we expect these configurations to be more
sensitive to the effect of the $\Delta^{++}$ MDM\footnote{One may also
chose to fit the whole set of UCLA data to  quote an average value for the
MDM (see for example Ref. \cite{Lin}). However, it is reasonable to
expect that the set of data that is most sensitive to the effects of the
MDM probes the underlying dynamics in a more detailed way.}. Once we would
have obtained a value of $\mu_{\Delta}$ from those configurations, we
would compare the results of our model with experimental data for other
geometrical configurations of photons.

   The fits of our model to the experimental data of Ref. \cite{ucla}
yield
the determinations of the magnetic dipole moment shown in Table 3, for the
three photon geometries and for two different energies of incident
pions: $T_{lab}= 269$ and $298$ MeV. The fitted curves to
$d\sigma/d\omega_{\gamma}d\Omega_{\pi} d\Omega_{\gamma}$ (Eq.\rf{6})
are represented
with solid lines in Figures 7 (for $T_{lab}=269$ MeV) and 8 (for
$T_{lab}=298$ MeV). In Figure 7 we have also included the curve (dashed
line)  corresponding to $\kappa_{\Delta}=1$, in order to visualize
the sensitivity of the differential cross section to the $\Delta^{++}$
MDM for those particular geometrical configurations. Although the
 experimental data are rather scarce, Fig. 7 clearly indicates
that the photon spectrum for the G1, G4 and G7 geometries is indeed
sensitive to the effect of the $\Delta^{++}$ MDM.

In order to compare
the sensitivity of different photon configurations, we have
displayed in Figure 9
the best fit  for the G7 configuration ($\kappa_\Delta$ = 3.27$\pm$0.76)
together with the cross section
evaluated at the same value of $\kappa_\Delta$ for the
configurations G11=$ (160^o,180^o)$ and
G14=$(103^o, 180^o$), for incident pions of
energy $T_{lab}=269$ MeV
\footnote{It is important to stress that a direct fit for these
configurations is not significant due to the small sensitivity of this
observable upon
$\kappa_\Delta$} . We observe that the curve corresponding to
$\kappa_{\Delta}=1$ (dashed line) is indistinguishable from the ones
corresponding to value $\kappa_{\Delta}= 3.27$ (full line), for
the G11 and G14 geometries.

  The different determinations of $\kappa_{\Delta}$ shown in
Table 3 are consistent among themselves. Therefore, we consider
meaningful to quote a weighted average over the six different fits. If we
express the weighted average in units of nuclear magnetons we obtain:
\begin{equation}
\mu_{\Delta} = 2(1+\kappa_{\Delta}) \frac{m_p}{m_{\Delta}} \left(
\frac{e}{2m_p} \right) =\left( 6.14 \pm 0.51 \right) \frac{e}{2 m_p}\ .
\label{15}
\end{equation}
In the Appendix we give an estimate of the effects of the MDM due to the
non-unitarity of the elastic scattering amplitude.

  Our result in Eq. (15) is compared in Table 4 with some of the
experimental determinations and theoretical calculations of the
$\Delta^{++}$ magnetic dipole moment that are available in the literature.
  Eq. (15) is compatible with the prediction $\mu_{\Delta}
=5.58 (e/2m_p)$ obtained in the SU(6) spin-flavor symmetry of the quark
model \cite{blp} and with the result obtained from the QCD and pion
corrections in the chiral bag-model \cite{krivo}; however our result is
somewhat larger than the prediction obtained
from the bag-model corrections to the SU(6) quark model, $\mu_{\Delta}=
(4.41 \sim 4.89)(e/2m_p)$ \cite{brv}.
Let us mention that a recent calculation based on a phenomenological quark
model that includes non-static effects
of pion exchange and orbital excitation \cite{Fra}, predicts the value
$\mu_{\Delta} = 6.17(e/2m_p)$ which fully agrees with our result. On
another hand,  Eq. (15) is also consistent with some previous
determinations of the MDM, for example with: $ \mu_{\Delta} =(5.6\sim
7.5) (e/2m_p)$ from Ref. \cite{wittman}, $(5.6 \pm 2.1)(e/2m_p)$
from Ref. \cite{pt} and $(6.9 \sim 9.8)(e/2m_p)$ from \cite{Hell}.
 However, the central value in Eq. (15) disagrees with some other
available determinations, namely:   $\mu_{\Delta} = (3.6 \pm
2)(e/2m_p)$ from \cite{Musa} obtained by using a relativistic model,  and
the result of Ref. \cite{Lin}, $(3.7 \sim 4.9)(e/2m_p)$, which is
obtained using a soft-photon approximation.

  Using the average value of $\mu_{\Delta}$ quoted in Eq. (15), we can
compare the results of our model with other angular configurations of
photon emission (labeled G1-G19 in Ref. \cite{ucla}). The $\chi^2$ values
obtained by comparing our prediction for
$d\sigma/d\omega_{\gamma}d\Omega_{\pi}d\Omega_{\gamma}$ and the
corresponding experimental values for the 18 detection angles, are shown
in Tables 5 and 6, respectively, for incident pions energies $T_{lab}=269$
MeV and $T_{lab}=298$ MeV. For illustrative purposes, in Figures 10 and
11 we have compared the
theoretical prediction based on Eq. (15) to experimental data of Ref.
\cite{ucla} for the special configurations G10-G15. According to the
$\chi^2$ values shown in Tables 5 and 6, we can conclude that our model
satisfactorily describes a large set of data of Ref. \cite{ucla}.

\bigskip
\bigskip
\begin{flushleft}
{\large \bf 5. Conclusions}
\end{flushleft}
\bigskip

  In the present paper we have used electromagnetic gauge invariance and
invariance under contact transformations as guiding principles to provide
a determination of the $\Delta^{++}$ magnetic dipole moment within a full
dynamical model for
the low energy interactions of this resonance. Our description of elastic
and radiative $\pi^+p$ scattering consistently incorporate the finite
width effects of the $\Delta^{++}$. The interaction vertices involving
this resonance do not require the inclusion of {\it ad hoc} form factors:
vertices correspond to structureless interactions as expected in the low
energy regime.  Resonant and background contributions to the
elastic and radiative $\pi^+p$ scattering amplitudes are cleanly
separated, as demanded by general principles of $S$-matrix theory.

  The relevant parameters of the $\Delta^{++}$ resonance (other than
the magnetic dipole moment) are fixed from the total cross section of
elastic $\pi^+p$ scattering.
The values obtained for the mass and width of
the $\Delta^{++}$ are very close to the ones obtained for these pole
parameters in a model-independent analysis of the $\pi^+p$ total cross
section based on the $S$-matrix theory \cite{bernicha}.
 The prediction for the differential cross section of elastic scattering
is found to be in very good agreement with available experimental data.
This consistency
check assures that our dynamical model gives a good description of
background contributions to elastic scattering as provided by the
$\Delta^0,\ N,\ \rho$ and $\sigma$ intermediate states.

  The magnetic dipole moment of the $\Delta^{++}$ turns out to be the only
adjustable parameter in radiative $\pi^+p$ scattering. Using the most
sensitive configurations of emitted photons, we have fitted the available
data of $d\sigma/d\omega_{\gamma}d\Omega_{\pi} d\Omega_{\gamma}$ in the
low-energy photon regime. We have found:
\[
\mu_{\Delta} = \left( 6.14 \pm 0.51 \right) \frac{e}{2 m_p}\ .
\]
Our result in consistent with predictions based on the simplest
version of SU(6) the quark model \cite{blp} and with a recent
quark model calculation that includes non-static states associated to pion
exchange and orbital excitations \cite{Fra}. The effects associated to the
non-unitarity of the $\pi^+ p$ scattering amplitude would change 
the value quoted above by a 2\%, as it is explained in the Appendix.

  With the value (Eq. (15)) extracted from the most sensitive angular
configurations in radiative $\pi^+p$ scattering, we can reproduce very
well a wider set of experimental data of Ref. \cite{ucla}. This conclusion
is based in the low $\chi^2$/dof values given in Tables 5 and 6, and the
illustrative plots shown in Figures 10 and 11.

{\large Acknowledgements}: The work of A. M. was  supported by
Conacyt (M\'exico). G.L.C. acknowledges partial support from Conacyt,
under contracts 32429-E and ICM-W-8015.

\newpage

\renewcommand{\theequation}{\thesection.\arabic{equation}} 
\appendix 
\section*{Appendix}
\setcounter{section}{1}
\setcounter{subsection}{1} 
\setcounter{equation}{0}

  In the model advocated in this paper, the amplitude for elastic
scattering
given in section 2 is not unitary. We
can easily check this by computing the imaginary part (dotted line in
Fig. 12) and the squared modulus (solid line in Fig. 12) 
of the elastic amplitude, Eq. (4). The lack of unitarity can be traced
back to the several approximations involved in obtaining Eq. (4), for
example: (a) the
$\Delta^{++}$ contributions include resummation of self-energy
corrections, while the remaining contributions are taken at the
tree-level;
(b) the decay width in the $\Delta^{++}$ propagator is taken as a
constant; (c) higher order corrections due to final state rescattering
have not been included.
In this Appendix we will not unitarize our scattering amplitude. Instead,
we will try to estimate the uncertainties in the free parameters of
elastic scattering processes associated to the lack of unitarity and then
we study how they would affect the determination of the $\Delta^{++}$ MDM.

  Within our model for $\pi^+p$ elastic scattering, the $\Delta^{++}$
($s$-channel) resonance plays the dominant role, while the $\Delta^0, \ n$
(in the crossed-channel) and the $\rho^0,\ \sigma$ ($t$-channel)
states give a subleading contribution. We call these contributions the
{\it resonant} and {\it background} terms, respectively. As is well known,
the singularity in the $s$-channel at $\sqrt{s}=m_{\Delta^{++}}$ is
avoided by 
using the Dyson resummation of self-energy graphs of the $\Delta^{++}$
resonance. Following Refs. \cite{fls}, we resum only the absorptive pieces
of these corrections and obtain an {\it improved} form of the Born
amplitude for the $\Delta^{++}$ contribution:
\begin{equation}
{\cal A}_{\Delta^{++}} = i\left(\frac{f_{\Delta \pi N}}{m_{\pi}} \right)
\frac{ \bar{u}(p') q'^{\mu} T_{\mu\nu} q^{\nu}
u(p)}{s-m_{\Delta}^2+i\sqrt{s}\Gamma_{\Delta}(s)}\ \label{A1},
\end{equation}
where we have defined the off-shell decay width of the $\Delta^{++}$
resonance (in its center-of-mass frame):
\begin{equation}
\sqrt{s}\Gamma_{\Delta}(s) = 
   =\left( \frac{f^2_{\Delta \pi N }}{4 \pi} \right) \times
     \frac{(\sqrt{s}+m_p)^2-m_{\pi}^2}{6 m_{\pi}^2 \sqrt{s} }\times
     |\vec{q}|^3 \label{A2}
\end{equation}
This energy-dependent width is computed from the interaction
Lagrangian (2) using the dominant decay mode
$\Delta^{++} \rightarrow p\pi^+$. The complex tensor $T_{\mu \nu}$ also
contains pieces of the absorptive corrections (see Refs. \cite{fls,bls}
for the case of spin-1 particles).

  Now, we can perform a Laurent expansion of ${\cal
A}_{\Delta^{++}}$ around the pole position $s_p\equiv
m_{\Delta}^2-im_{\Delta}\Gamma_{\Delta}$ \cite{z0}. By retaining
leading order terms, we obtain:
\begin{equation}
{\cal A}_{\Delta^{++}} = {\cal M}_{\Delta^{++}} + \widetilde{\cal A}(s)\ ,
\label{A3}
\end{equation}
where ${\cal M}_{\Delta^{++}}$ (given by Eqs. (5)-(6)) has the same form
as the Born amplitude but where $m_{\Delta}^2 \rightarrow
m_{\Delta}^2-im_{\Delta}\Gamma_{\Delta}$, and $\widetilde{\cal A}(s)$ is a
regular function in the resonance region \cite{z0}.

   From our experience with spin-1 resonances (see M. Beuthe et al.
\cite{fls} and Ref. \cite{bls}), we can state that Eq. (A1) would reduce
to the first term in Eq. (A3) in the limit where the pions and protons
involved
in the loop corrections ($\Delta^{++} \rightarrow \pi^+ p \rightarrow
\Delta^{++}$)  are massless. Furthermore, when the full $\Delta^{++}$
propagators in Eqs. (A1) and in ${\cal M}_{\Delta^{++}}$ are combined with
the corresponding $\gamma \Delta^{++} \Delta^{++}$ vertices satisfy a Ward
identity that assures gauge-invariance in the radiative $\pi^+ p$
scattering (see Eq. (14) in Ref. \cite{elamiri}).

   Now, the total amplitude for elastic scattering is given by:
\begin{eqnarray}
{\cal M}_T &=& {\cal M}_{\Delta^{++}} + \widetilde{\cal A}(s) + B(s)
\nonumber \\ &\equiv& {\cal M} + \widetilde{\cal A}(s)
+ \widetilde{\cal B}(s) \label{A4}
\end{eqnarray}
where
\begin{eqnarray}
B(s) &=& \sum_{i=\Delta^0, n,\rho,\sigma} {\cal M}_{i}
+ \widetilde{\cal B}(s)\nonumber,
\end{eqnarray}
contains the tree-level amplitudes with $\Delta^0,\ n,\ \rho$
and $\sigma$ intermediate states, and the rescattering corrections
$\widetilde{\cal B}(s)$.
 We expect that these last effects would not generate a resonant-like
structure and, consequently, can be absorbed
into background contributions by replacing $\widetilde{\cal A}(s) 
+ \widetilde{\cal B}(s)
\rightarrow {\cal A}_{eff}(s)$ in Eq. (A4). As it is clear, in the
analysis of the present
paper, we have neglected the term ${\cal A}_{eff}(s)$. In the following, we
will argue how this term can help to restore unitarity in elastic $\pi^+
p$ scattering in our model.

  Since ${\cal A}_{eff}(s)$ is a slowly varying function of $s$ in the
resonant region (see Fig. 3), and since we expect ${\cal A}_{eff}(s) <<
{\cal
M}$, we can write as a first approximation:
\begin{equation}
{\cal M}_T \approx {\cal M} e^{-i\delta} \label{A5}\ ,
\end{equation}
where the global phase $\delta$ is close to zero, given the above
considerations. Let us assume that $\cal M$ is not unitary
(as it is the case in the present work), namely it does {\it not}
 satisfy:
\be
-2 Im {\cal M} = |{\cal M}|^2\ \label{A6}.
\ee
The left and right members of Eq. (A6) correspond to the
dotted and solid lines in Figure 12.  
We have checked numerically that the unitarity condition $-2 Im {\cal M}_T
= |{\cal M}_T|^2$ is indeed satisfied when $\delta \approx 10^0$
(corresponding to the dashed-dotted line in Figure 12),
which is in agreement with the expectation that ${\cal A}_{eff} <<
{\cal M}$. Note that the solid line in Figure 12 corresponds to the best
fit obtained in our work.

   In order to check this hypothesis further, we have performed a new fit
to experimental data using Eq. (A4), with a linear parametrization ${\cal
A}_{eff} = as+b$. All the parameters change within error bars quoted in
Table 1 except the mass and width of the $\Delta^{++}$ which become
$m_{\Delta^{++}} = 1211.7$ MeV and $\Gamma_{\Delta}=92.2$ MeV (to be
compared with the last row in Table 1). These
values are even closer to the ones obtained in the model-independent
analysis of Ref. \cite{bernicha}, with an amplitude that satisfies
unitarity.

   Now, we can estimate the effects of unitarizing the elastic amplitude
of our model on the determination of the $\Delta^{++}$ magnetic dipole
moment from radiative $\pi^+ p$ scattering. We have performed new fits to
the set of data on radiative pion-proton scattering that is sensitive to
the effect of the $\Delta^{++}$ MDM. If we now set $m_{\Delta}=1212$ MeV
and $\Gamma_{\Delta} =95$ MeV, we obtain the results quoted in Table 7.
When we compare Tables 3 and 7, we observe that the $\Delta^{++}$
MDM is almost insensitive to the precise values of the mass and
width of this resonance. The only change observed is an increase in the 
error bars of the MDM. Note that to obtain the results in Table 7 we have
allowed a conservative large increase in the width of the $\Delta^{++}$
resonance (with respect to the values in Table 1). With the values
reported in Table 7, we obtain the weighted average of the $\Delta^{++}$
MDM, which now becomes:
\be
\mu_{\Delta^{++}} = (6.01 \pm 0.61) \frac{e}{2m_p}\ ,
\ee
which is consistent with the value shown in Eq. (15). We conclude that
unitarity only affects the determination of the $\Delta^{++}$ MDM within
the error bars already determined by our non-unitary approximation.

\newpage
\bigskip
\bigskip
\bc
{\bf \large TABLE CAPTIONS}
\ec

\bigskip

Table 1: Yields from the fits  to the total cross section of the elastic
$\pi^+p$ scattering by including the different intermediate states
($g_{\sigma} \equiv g_{\sigma \pi\pi}g_{\sigma NN}$).
\bigskip

Table 2: Fits to the cross section of the elastic $\pi^+p$
scattering including $\Delta^{++, 0},N,\rho$,and$\sigma$
degrees of freedom for different values of $m_\sigma$.
\bigskip

Table 3: Fits to the differential cross section of radiative
$\pi^+p$ scattering, for three different geometries of photon
angle emission and two different energies of incident pions.
\bigskip

Table 4: Some of the experimental determinations and theoretical
calculations of the $\Delta^{++}$ magnetic dipole moment.
\bigskip

Table 5: $\chi^2$/(\# data) values obtained from the comparison between
our model (based on Eq. (15)) and experimental data, for the
differential cross section of radiative $\pi^+p$ scattering at
$T_{lab}$=269 MeV. G1-G19 label the geometrical configurations as in Ref.
\cite{ucla}.
\bigskip

Table 6: Same as Table 5 for $T_{lab}$=298 MeV.
\bigskip

Table 7: Same as Table 3, with unitarity effects included.

\newpage
\bigskip
\bigskip
\bc
{\bf \large FIGURE CAPTIONS}
\ec

\bigskip

Figure 1: Feynman graphs corresponding to the different contributions to
the elastic $\pi^+ p$ scattering amplitude.
\bigskip

Figure 2: Feynman graphs for the radiative $\pi^+ p$ amplitude:
$s$-, crossed-, and $t$-channels correspond to $\Delta^{++}$, $(\Delta^0,
N)$ and $(\rho,\sigma)$ contributions in intermediate states,
respectively.
\bigskip

Figure 3: Total cross section of elastic $\pi^+p$ scattering as a function
of incident $\pi^+$ kinetic energy.
\bigskip

Figure 4: Differential cross section of elastic  $\pi^+ p$ scattering.
Circles, triangles and squares denote, respectively, experimental data
from Refs.\cite{Bussey, Sadler, Gordeev}. The curves (with same convention
as in Figure 1) denote our prediction for
$T_{lab} = 263.7$ MeV  (upper box) and $T_{lab} = 291.4$ MeV (lower box).
\bigskip

Figure 5: Elastic $\pi^+ p$ total cross section as a function of incident
$\pi^+$ kinetic energy obtained for three different  masses of the
$\sigma$ meson.
\bigskip

Figure 6: Differential cross section of elastic  $\pi^+ p$ scattering,
for $T_{lab} = 263.7$ MeV  (upper box) and $T_{lab} = 291.4$ MeV (lower
box) and $m_{\sigma} = 450, 650$ and 850 MeV. Line conventions are those
of Figure 5 and experimental results correspond to Refs.\cite{Bussey,
Sadler, Gordeev}.

\bigskip

Figure 7: Differential cross section of radiative $\pi^+ p$
 scattering for $T_{lab} = 269$ MeV. The  G1, G4 and G7 geometries are
defined in Table 2. The solid line corresponds to the best fit and the
dashed line to $\kappa_{\Delta}=1$.
\bigskip

Figure 8: Same as in Figure 7 when $T_{lab}= 298$ MeV.
\bigskip

Figure 9:  Same as in Figure 7 for the G7, G11, and G14 configurations,
but
the full lines correspond in all cases to values of $\kappa_\Delta$
obtained from the best fitting for G7 case (see section 4.2).
\bigskip

Figure 10: Differential cross section of radiative $\pi^+ p$ scattering
for $T_{lab} = 269$ MeV and the G10-G15 configurations \cite{ucla}. The
solid lines correspond to the prediction based on Eq. (15).

Figure 11: Same as in Figure 10 when $T_{lab}=298$ MeV \cite{ucla}.

Figure 12: Total cross section of elastic $\pi^+p$ scattering as a function
of the incident $\pi^+$ kinetic energy . Full lines correspond to the
best fit in Figure 1. The dotted-dashed (dotted) line represents the
imaginary part of the amplitude that satisfies (does not satisfy)
unitarity. See the appendix for explanations. 

\bigskip

\newpage
\begin{center}
\begin{tabular}{|c|c|c|c|c|c|}
\hline\hline
Interm. state & $f_{\Delta N \pi}^2/4\pi\ $ &
$m_{\Delta}$ (MeV) &
$\Gamma_{\Delta}$ (MeV) & $g_{\sigma}/4\pi$ & $\chi^2$/dof \\
\hline
$\Delta^{++, 0}$ & 0.281$\pm$0.001 & 1201.7$\pm$0.2 & 69.8$\pm$0.2 &
-- & 121.1\\
$\Delta^{++, 0}, N$ & 0.331$\pm$0.003 & 1208.6$\pm$0.2 & 87.5$\pm$0.3 &
-- & 17.6 \\
$\Delta^{++, 0}, N, \rho$ & 0.327$\pm$0.001 & 1207.4$\pm$0.2 &
85.6$\pm$0.3 & -- & 15.6 \\
$\Delta^{++, 0},N,\rho, \sigma$ & 0.317$\pm$0.003 & 1211.2$\pm$0.4 &
88.2$\pm$0.4 & 1.50$\pm$0.12 & 10.5 \\
\hline
\end{tabular}
\end{center}
\bigskip
\centerline{\bf \large Table 1}
\bigskip
\bigskip
\bigskip
\bigskip

\begin{center}
\begin{tabular}{|c|c|c|c|c|}
\hline\hline
$m_\sigma [MeV]$ & $f_{\Delta N
\pi}^2/4\pi\ $ &
$m_{\Delta}$ (MeV) &
$\Gamma_{\Delta}$ (MeV) & $g_{\sigma \pi\pi}g_{\sigma NN}/4\pi$ \\
\hline
$450$ & 0.311$\pm$0.002 & 1211.3$\pm$0.3 & 86.4$\pm$0.3 &
1.00 $\pm$0.05\\
$650$ & 0.317$\pm$0.003 & 1211.2$\pm$0.4 &
88.2$\pm$0.4 &
1.50 $\pm$0.11 \\
$850$ & 0.318$\pm$0.003 & 1211.1$\pm$0.4 &
88.5$\pm$0.4 & 1.90 $\pm$0.18 \\
\hline
\end{tabular}
\end{center}
\bigskip

\centerline{\bf \large Table 2}
\bigskip
\bigskip
\bigskip
\bigskip

\begin{center}
\begin{tabular}{|c|c|c|c|c|c|}
\hline\hline
$T_{lab}$ (MeV)&Geometry &$\theta_{\gamma}$& $\phi_{\gamma}$ &
$\kappa_{\Delta}$ & $\chi^2$/dof  \\
\hline
& G7 & 120$^0$ & 0$^0$ & 3.27$\pm$0.76 & 1.99\\
269& G4 & 140$^0$ & 0$^0$ & 3.01$\pm$0.67 & 2.48\\
&G1& 160$^0$ & 0$^0$ & 2.74$\pm$0.87  & 1.73\\
\hline
&G7& 120$^0$ & 0$^0$ & 3.10$\pm$0.87 & 2.68\\
298 & G4 & 140$^0$ & 0$^0$ & 2.90$\pm$0.75 & 4.75\\
&G1 &160$^0$ & 0$^0$ & 2.61$\pm$1.00 & 1.47\\
\hline
\end{tabular}
\end{center}
\bigskip

\centerline{\bf \large Table 3}

\newpage
\bigskip
\bigskip
\bigskip
\bigskip

\begin{center}
\begin{tabular}{|c|c|c|c|}
\hline
Source & $\mu_{\Delta} \times e/2m_p$ & Experimental data & Method/model
\\
\hline
& Experimental determinations & \\
\hline
Ref. \cite{Musa} & 3.6$\pm$2.0 & Ref. \cite{Arman} & Relativistic \\
Ref. \cite{pt} & 5.64$\pm$2.13 & Ref. \cite{Leung} & Relativistic  \\
Ref. \cite{ucla} & 4.7$\sim$6.7 & Ref. \cite{ucla} & Model of Ref.
\cite{pt} and \\
& & & Soft-photon approx. (SPA) \\
Ref. \cite{Hell} & 6.9$\sim$9.8 & Ref. \cite{ucla} & MIT isobar of Ref.
\cite{Hell} \\
Ref. \cite{wittman} & 5.6$\sim$7.5 & Ref. \cite{ucla} & Relativistic \\
Ref. \cite{sin} & 4.58$\pm$0.33 & Ref. \cite{sin} & MIT isobar
\cite{Hell} \\
Ref. \cite{Lin} & 3.7$\sim$4.9 & Ref. \cite{ucla, sin} & Modified SPA \\
Present work & 6.14$\pm$0.51 & Ref. \cite{ucla} & Relativistic \\
\hline
& Theoretical calculations & \\
\hline
Ref. \cite{blp} & 5.58 & -- & SU(6) symmetry \\
Ref. \cite{bp} & 4.25 & -- & Mass corrections to SU(6) \\
Ref. \cite{brv} & 4.41$\sim$4.89 & -- & $\pi$-cloud corr. to SU(6) \\
Ref. \cite{krivo} & 6.54 & -- & QCD corr. in $\chi$ bag-model \\
Ref. \cite{Fra} & 6.17 & -- & Phenom. quark model \\
\hline
\end{tabular}
\end{center}
\bigskip

\centerline{\bf \large Table 4}

\newpage
\bigskip
\bigskip
\bigskip

\begin{center}
\begin{tabular}{|c|c|c|c|c|}
\hline
$T_{lab}$ (MeV)&Geometry &$\theta_{\gamma}$& $\phi_{\gamma}$
& $\chi^2$/(\# data) \\
\hline
& G1 & 160$^0$ & 0$^0$ & 0.8  \\
& G2 & 153.3$^0$ & 316.5$^0$ & 2.1 \\
& G3 & 139.5$^0$ & 295.2$^0$  & 2.6  \\
& G4 & 160$^0$ & 0$^0$  & 2.  \\
& G5 & 136$^0$ & 333.2$^0$  & 1.7  \\
& G6 & 128.3$^0$ & 311.5$^0$  & 8.7  \\
& G7 & 120$^0$ & 0$^0$  & 2.1  \\
& G8 & 118.4$^0$ & 339.4$^0$  & 6.1  \\
& G9 & 113.8$^0$ & 320$^0$  & 3.7  \\
269& G10 & 144$^0$ & 270$^0$  & 0.5 \\
& G11 & 160$^0$ & 180$^0$  & 2.3  \\
& G12 & 140$^0$ & 180$^0$  & 1.5  \\
& G13 & 120$^0$ & 180$^0$  & 0.7 \\
& G14 & 103$^0$ & 180$^0$  & 0.9  \\
& G15 & 102.2$^0$ & 200.5$^0$  & 0.40  \\
& G17 & 50.1$^0$ & 174.8$^0$  & 1.  \\
& G18 & 64.6$^0$ & 293.4$^0$  & 0.8  \\
& G19 & 59$^0$ & 270$^0$  & 0.3  \\
\hline
\end{tabular}
\end{center}
\bigskip

\centerline{\bf \large Table 5}

\newpage
\bigskip
\bigskip
\bigskip
\begin{center}
\begin{tabular}{|c|c|c|c|c|}
\hline
$T_{lab}$ (MeV)&Geometry &$\theta_{\gamma}$& $\phi_{\gamma}$
& $\chi^2$/(\# data) \\
\hline
& G1 & 160$^0$ & 0$^0$  & 0.4  \\
& G2 & 153.3$^0$ & 316.5$^0$  & 0.2 \\
& G3 & 139.5$^0$ & 295.2$^0$  & 0.9  \\
& G4 & 160$^0$ & 0$^0$  & 3.6  \\
& G5 & 136$^0$ & 333.2$^0$  & 2.8  \\
& G6 & 128.3$^0$ & 311.5$^0$  & 0.6  \\
& G7 & 120$^0$ & 0$^0$  & 2.4  \\
& G8 & 118.4$^0$ & 339.4$^0$  & 0.5  \\
& G9 & 113.8$^0$ & 320$^0$  & 2.7  \\
298& G10 & 144$^0$ & 270$^0$  & 1. \\
& G11 & 160$^0$ & 180$^0$  & 0.4  \\
& G12 & 140$^0$ & 180$^0$  & 4.9  \\
& G13 & 120$^0$ & 180$^0$  & 2.4 \\
& G14 & 103$^0$ & 180$^0$  & 1.  \\
& G15 & 102.2$^0$ & 200.5$^0$  & 0.5  \\
& G17 & 50.1$^0$ & 174.8$^0$  & 0.8  \\
& G18 & 64.6$^0$ & 293.4$^0$  & 1.1  \\
& G19 & 59$^0$ & 270$^0$  & 1.31  \\
\hline
\end{tabular}
\end{center}
\bigskip

\centerline{\bf \large Table 6}

\newpage
\bigskip
\bigskip
\bigskip
\bigskip
\begin{center}
\begin{tabular}{|c|c|c|c|c|c|}
\hline\hline
$T_{lab}$ (MeV)&Geometry &$\theta_{\gamma}$& $\phi_{\gamma}$ &
$\kappa_{\Delta}$ & $\chi^2$/dof  \\
\hline
& G7 & 120$^0$ & 0$^0$ & 3.27$\pm$1.03 & 1.46\\
269& G4 & 140$^0$ & 0$^0$ & 3.01$\pm$0.85 & 2.24\\
&G1& 160$^0$ & 0$^0$ & 2.74$\pm$1.04  & 1.00\\
\hline
&G7& 120$^0$ & 0$^0$ & 3.10$\pm$0.7 & 1.80 \\
298 & G4 & 140$^0$ & 0$^0$ & 2.90$\pm$1.0 & 3.57\\
&G1 &160$^0$ & 0$^0$ & 2.61$\pm$1.00 & 1.44\\
\hline
\end{tabular}
\end{center}
\bigskip

\centerline{\bf \large Table 7}

\newpage
\begin{figure}
\begin{center}
    \leavevmode
   \epsfxsize = 18cm
     \epsfysize = 20cm
    \epsffile{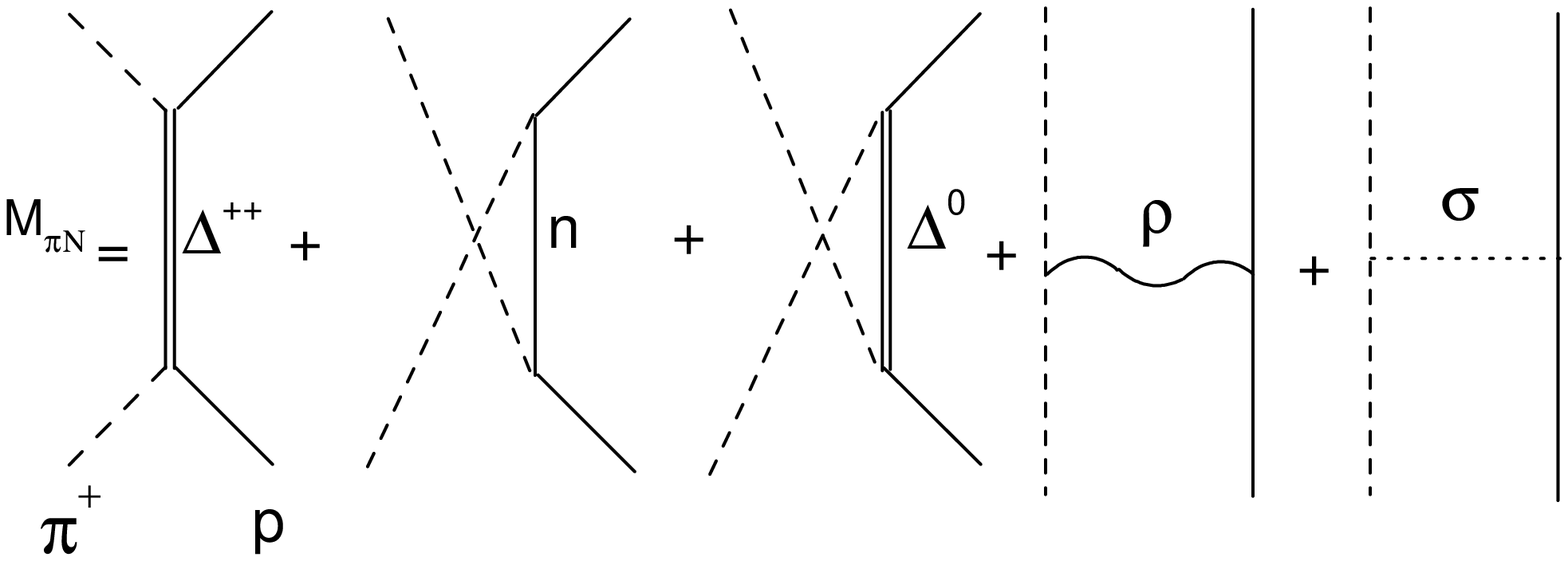}
   \end{center}
\vspace{-6.5cm}
\label{fig1}
\end{figure}
\centerline{\bf\large Figure 1}
\newpage
\begin{figure}
\begin{center}
    \leavevmode
   \epsfxsize = 15cm
     \epsfysize = 15cm
    \epsffile{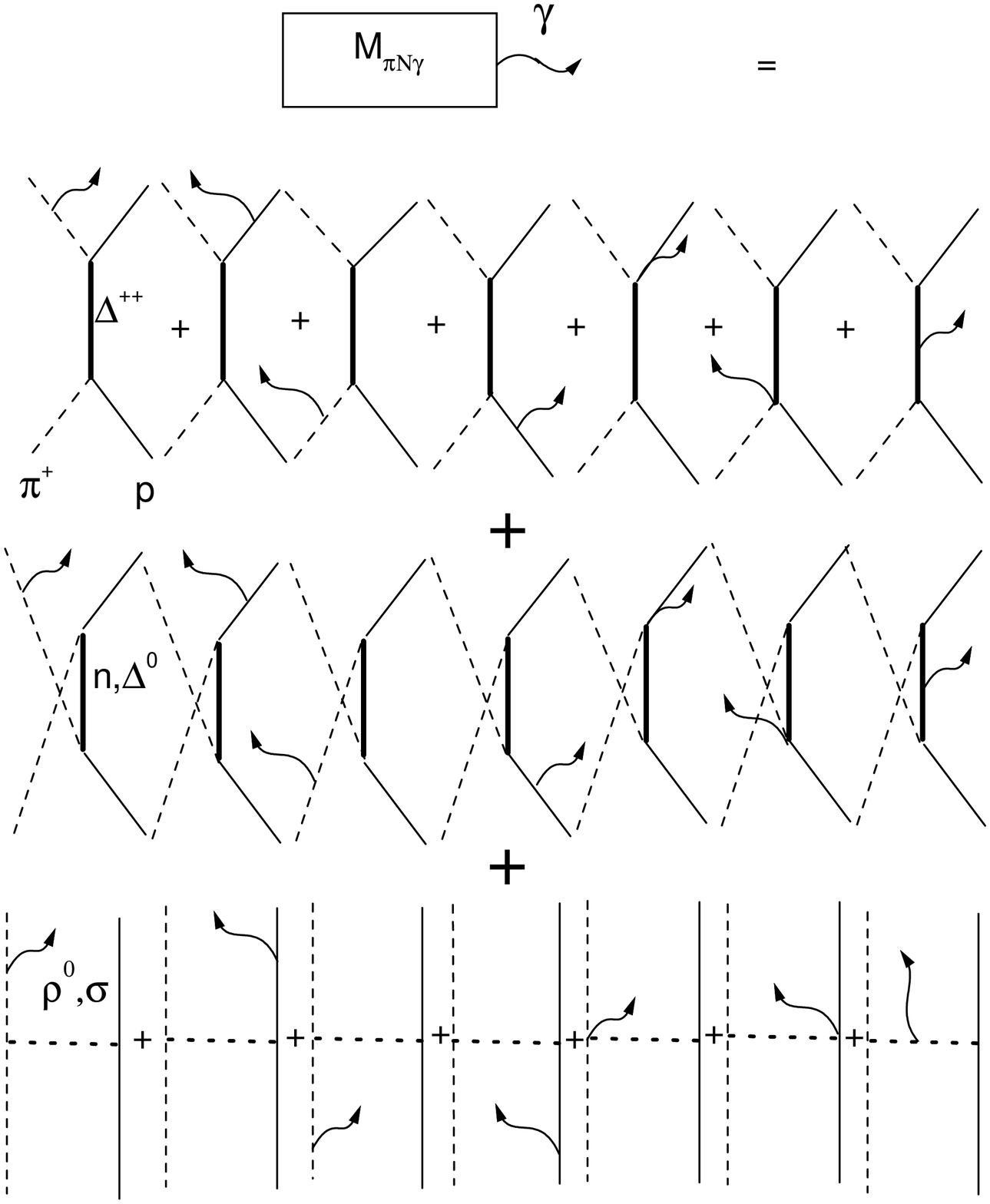}
   \end{center}
\vspace{-1.cm}
\label{fig2}
\end{figure}
\centerline{\bf\large Figure 2}
\newpage
\begin{figure}
\begin{center}
    \leavevmode
   \epsfxsize = 15cm
     \epsfysize = 16cm
\vspace{-2.5cm}
    \epsffile{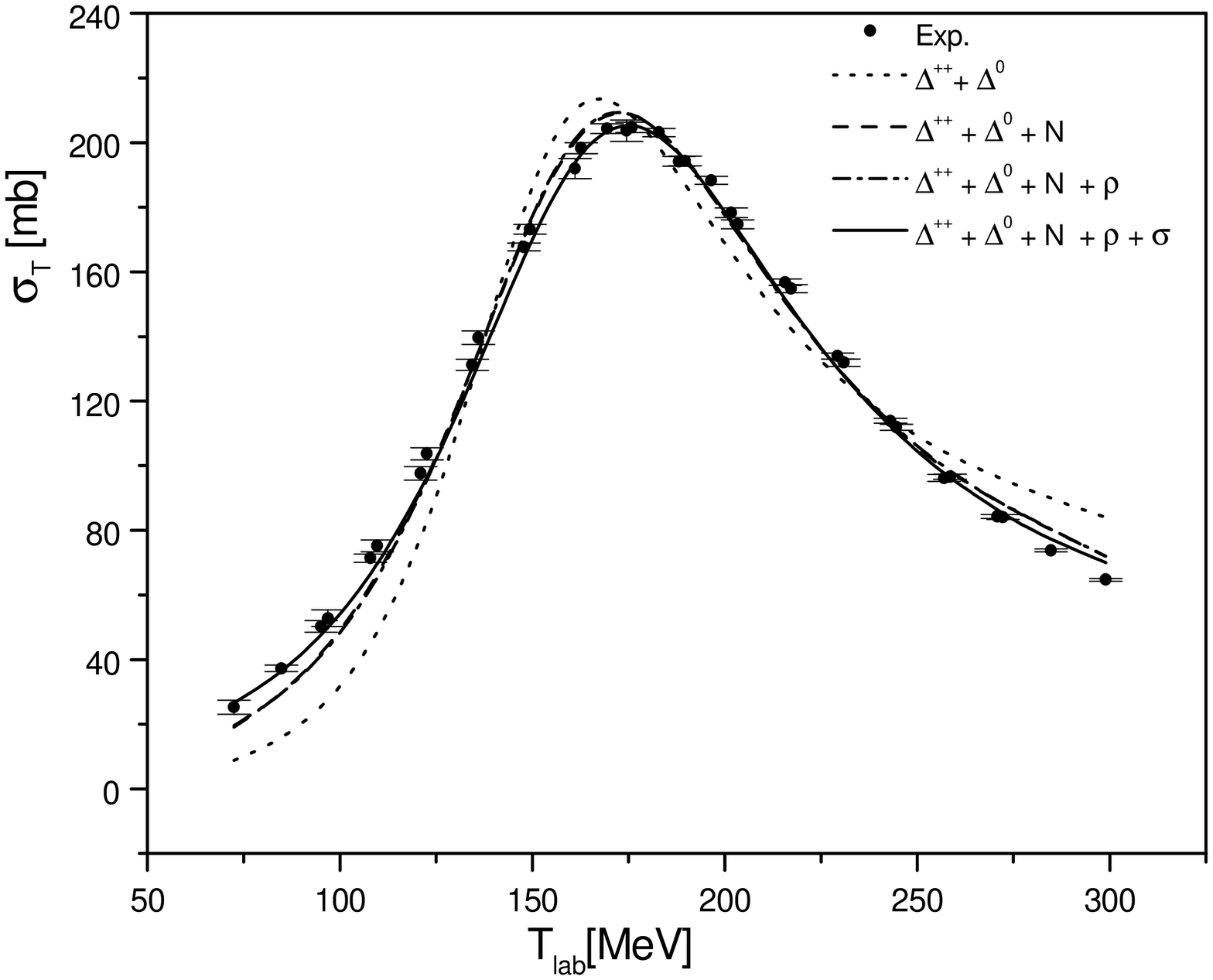}
   \end{center}
\vspace{-1.cm}
\end{figure}
\centerline{\bf\large Figure 3}
\newpage

\begin{figure}
\begin{center}
    \leavevmode
   \epsfxsize = 16cm
     \epsfysize = 17cm
\vspace{-2.5cm}
    \epsffile{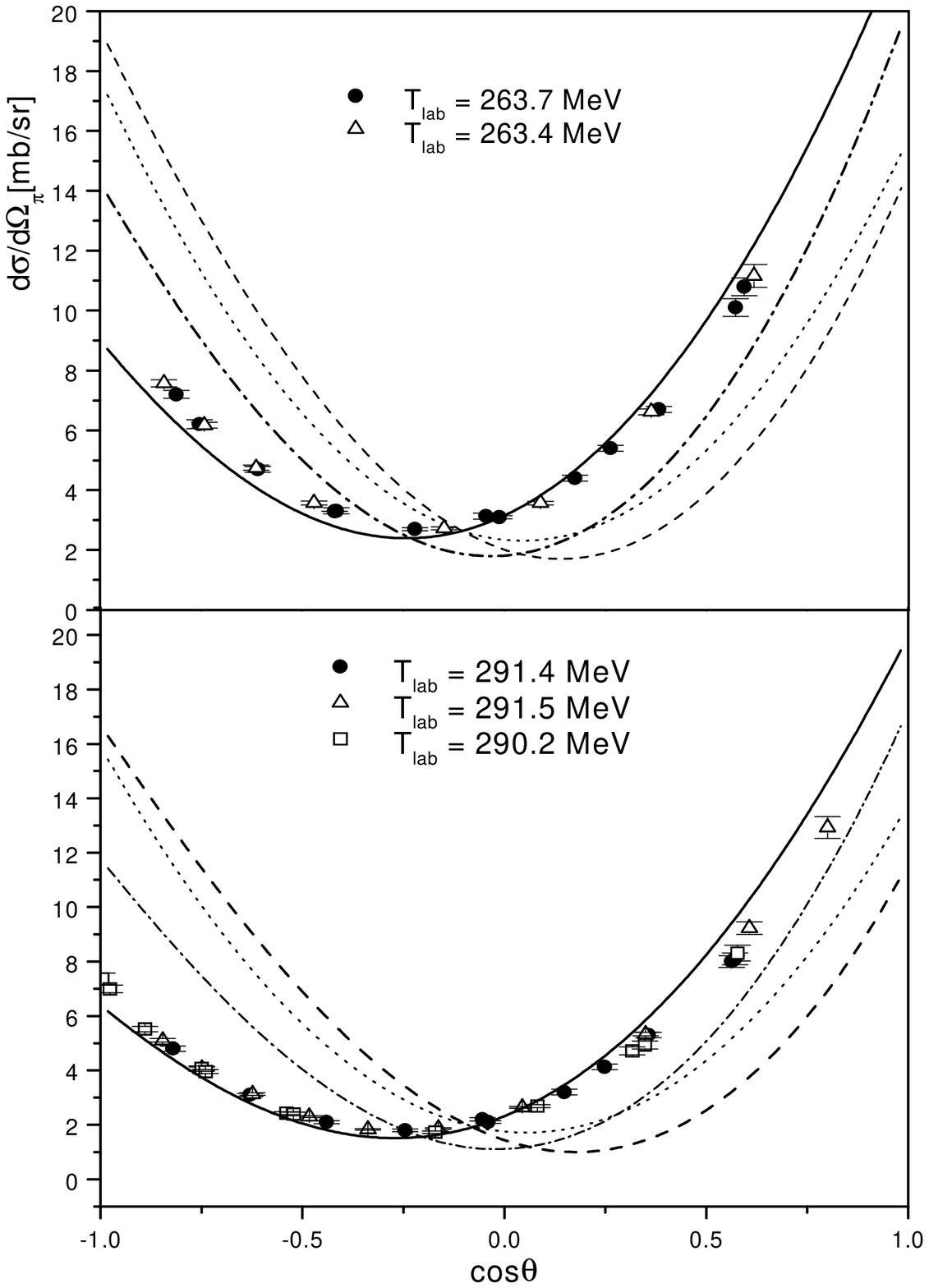}
   \end{center}
\vspace{-1.cm}
\end{figure}

\centerline{\bf\large Figure 4}
\newpage

\begin{figure}
\begin{center}
    \leavevmode
   \epsfxsize = 15cm
     \epsfysize = 16cm
    \epsffile{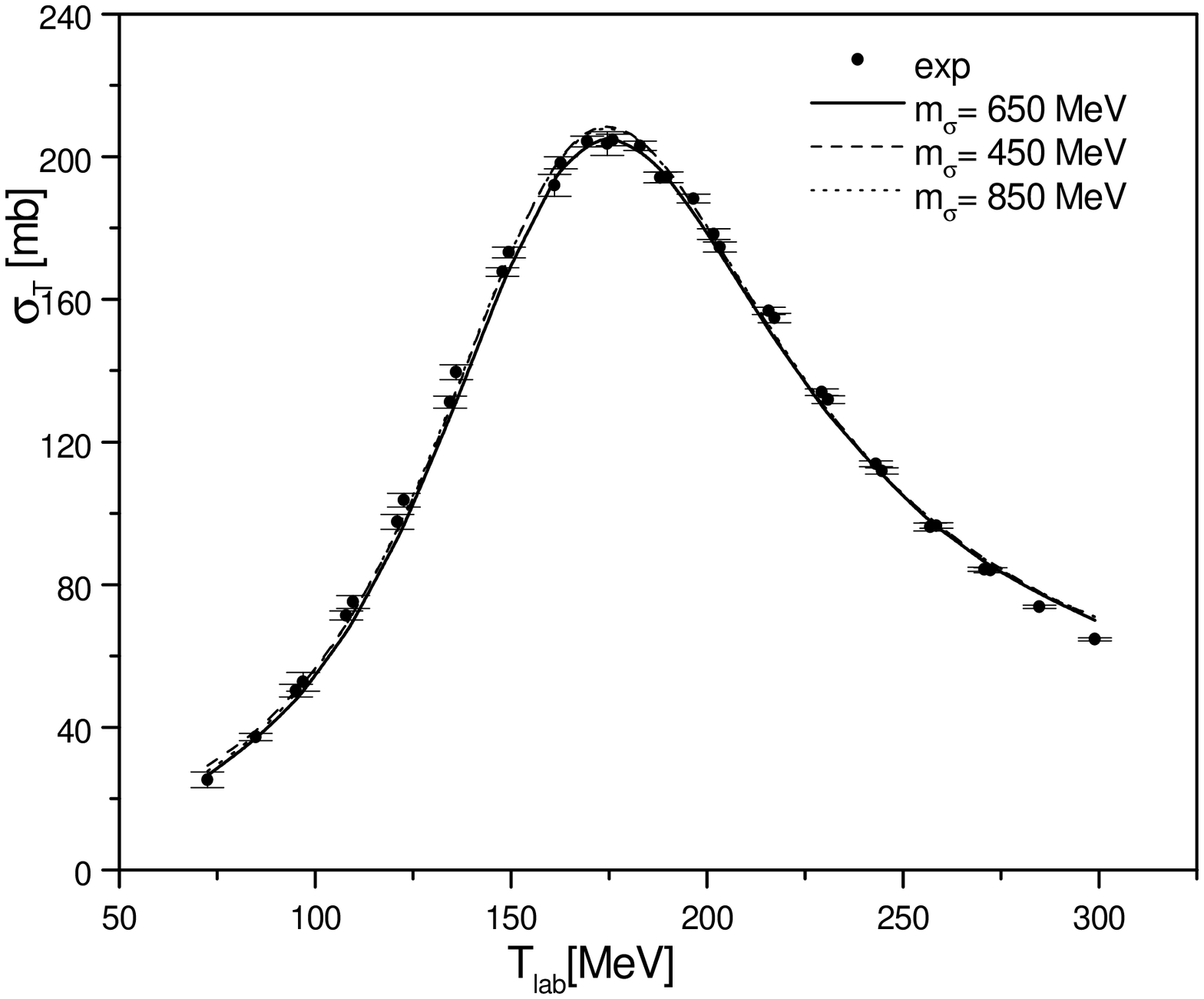}
   \end{center}
\vspace{-3.5cm}
\end{figure}
\centerline{\bf\large Figure 5}
\newpage
\vspace{-3cm}
\begin{figure}
\begin{center}
    \leavevmode
   \epsfxsize = 16cm
     \epsfysize = 17cm
    \epsffile{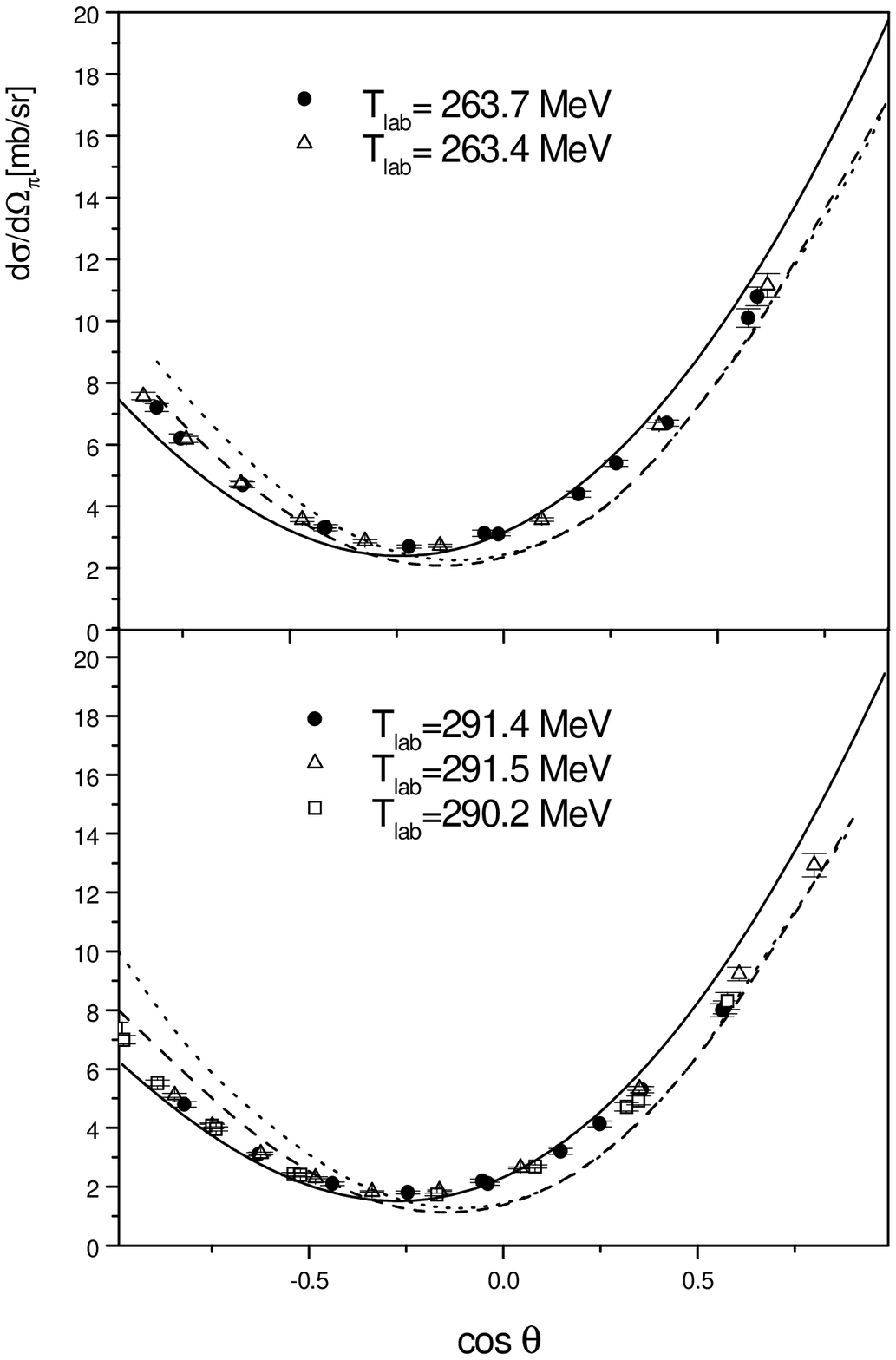}
   \end{center}
\vspace{-3.cm}
\end{figure}
\centerline{\bf\large Figure 6}
\newpage
\begin{figure}
\begin{center}
    \leavevmode
   \epsfxsize = 15cm
     \epsfysize = 16cm
    \epsffile{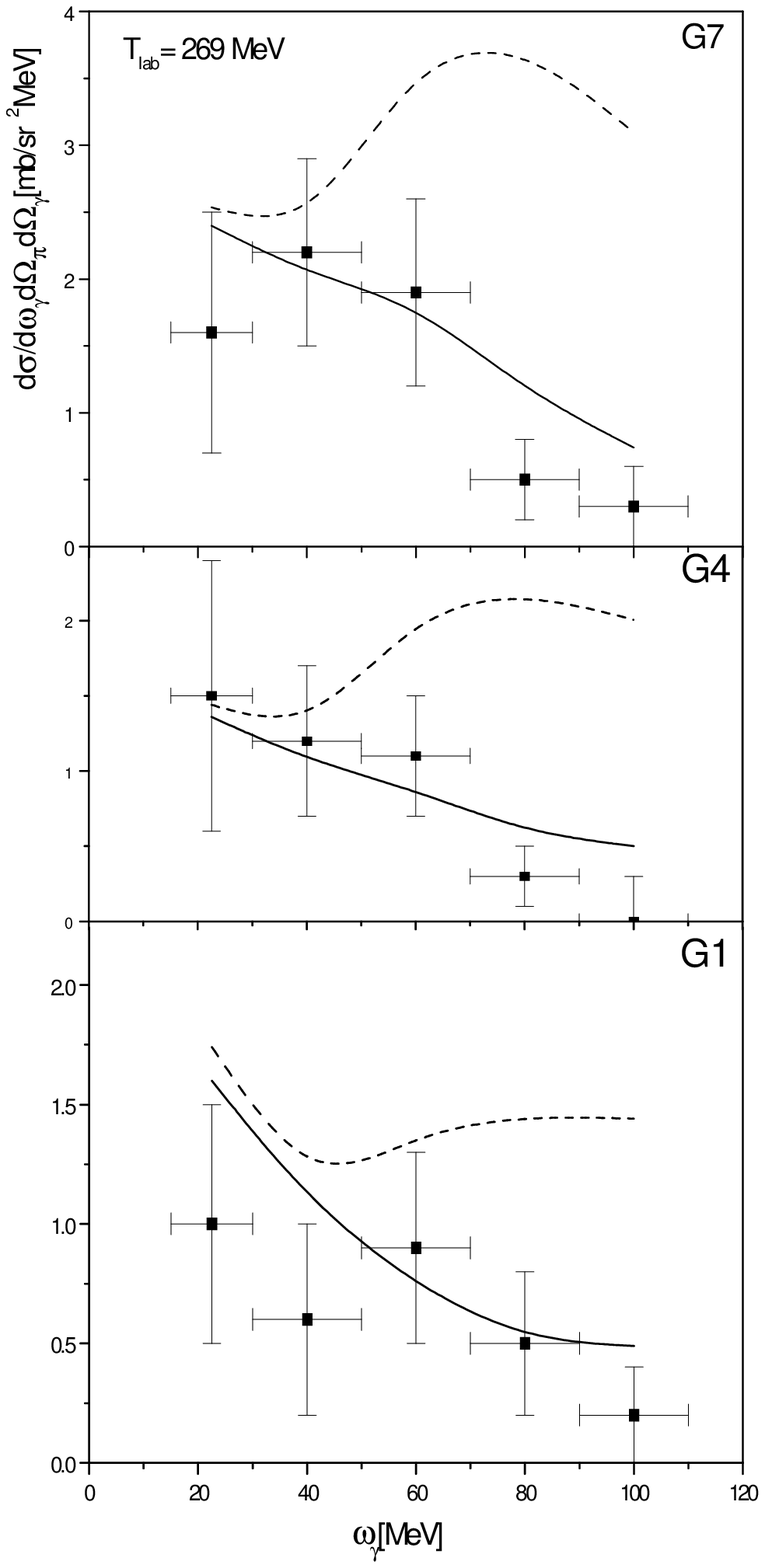}
   \end{center}
\vspace{-2.5cm}

\end{figure}
\centerline{\bf\large Figure 7}
\newpage

\begin{figure}
\begin{center}
    \leavevmode
   \epsfxsize = 17cm
     \epsfysize = 18cm
    \epsffile{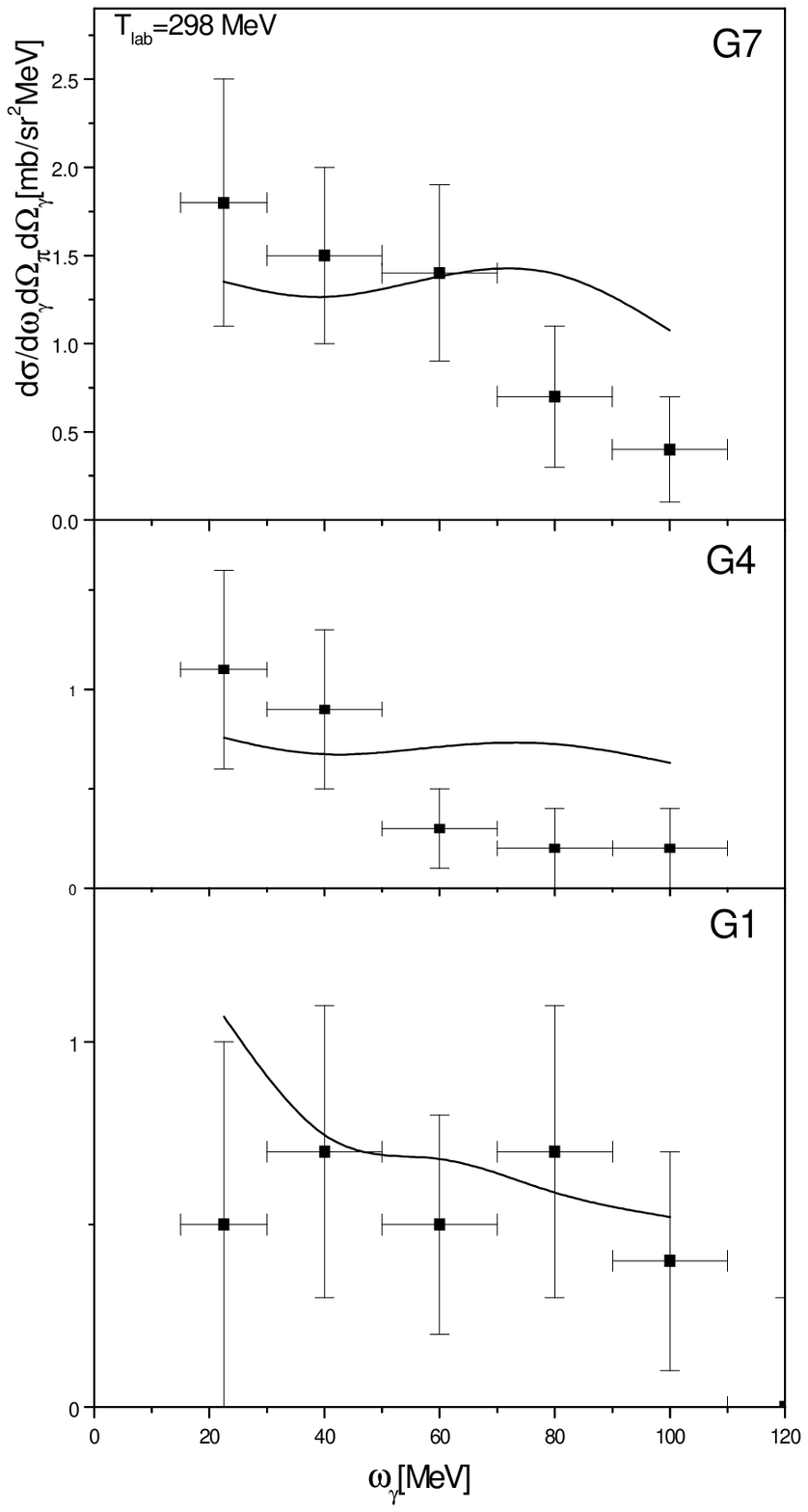}
   \end{center}
\vspace{-4.cm}
\end{figure}
\centerline{\bf\large Figure 8}
\newpage

\begin{figure}
\begin{center}
    \leavevmode
   \epsfxsize = 15cm
     \epsfysize = 17cm
    \epsffile{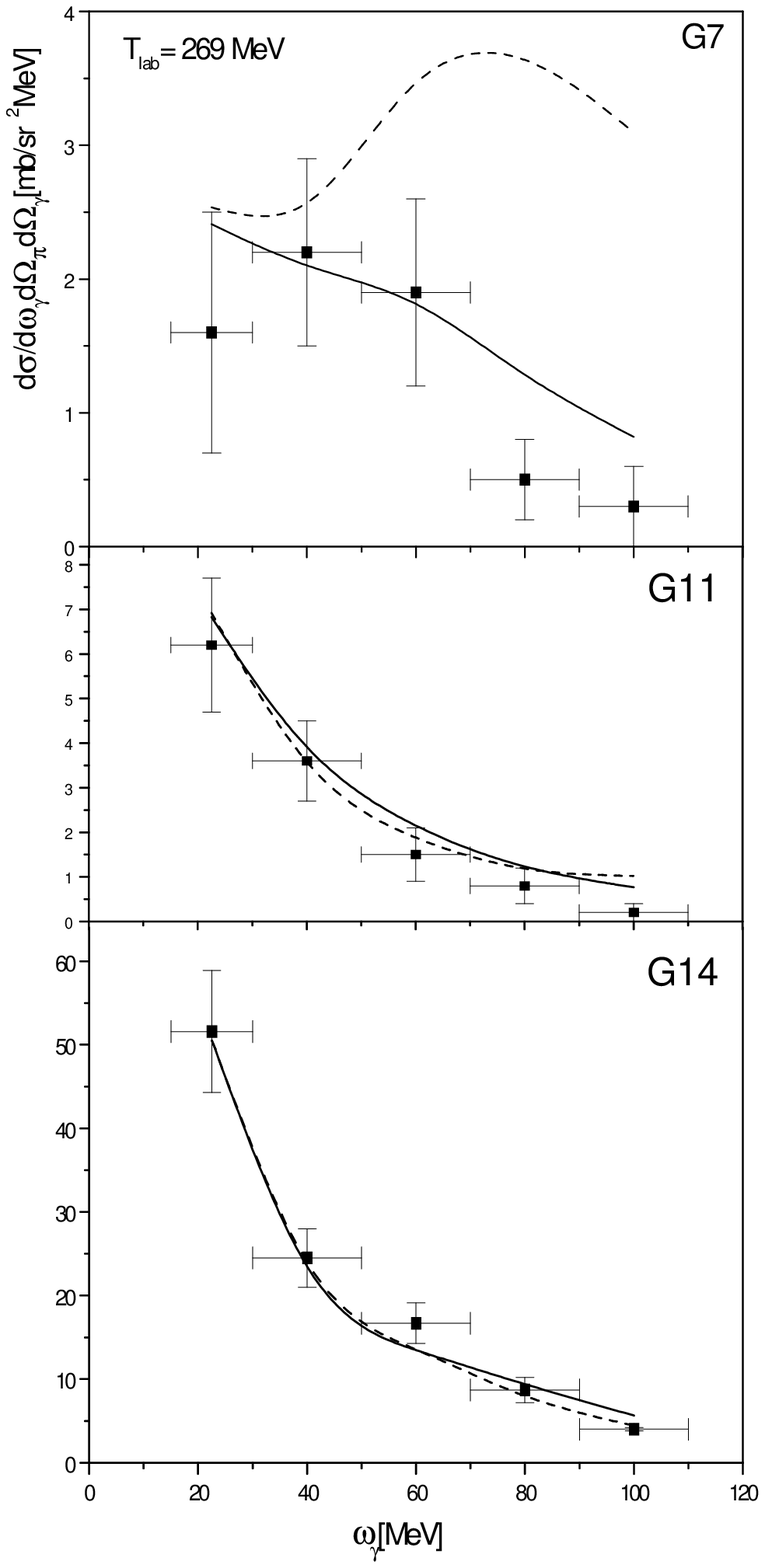}
   \end{center}
\vspace{-3.cm}
\end{figure}
\centerline{\bf\large Figure 9
}

\newpage
\begin{figure}
\begin{center}
    \leavevmode
   \epsfxsize = 15cm
     \epsfysize = 17cm
    \epsffile{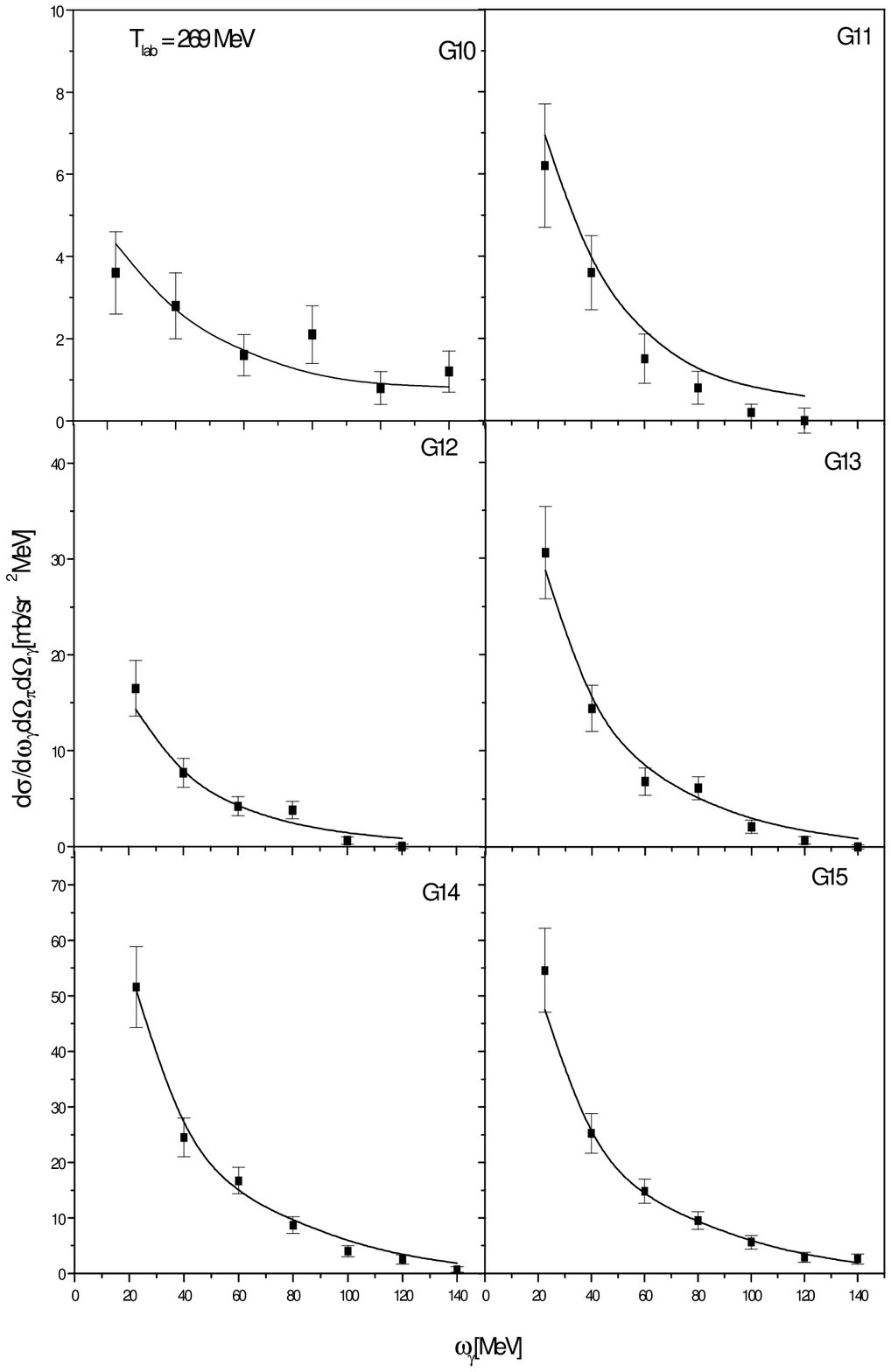}
   \end{center}
\vspace{-3.cm}
\end{figure}
\centerline{\bf\large Figure 10
}

\newpage
\begin{figure}
\begin{center}
    \leavevmode
   \epsfxsize = 15cm
     \epsfysize = 17cm
    \epsffile{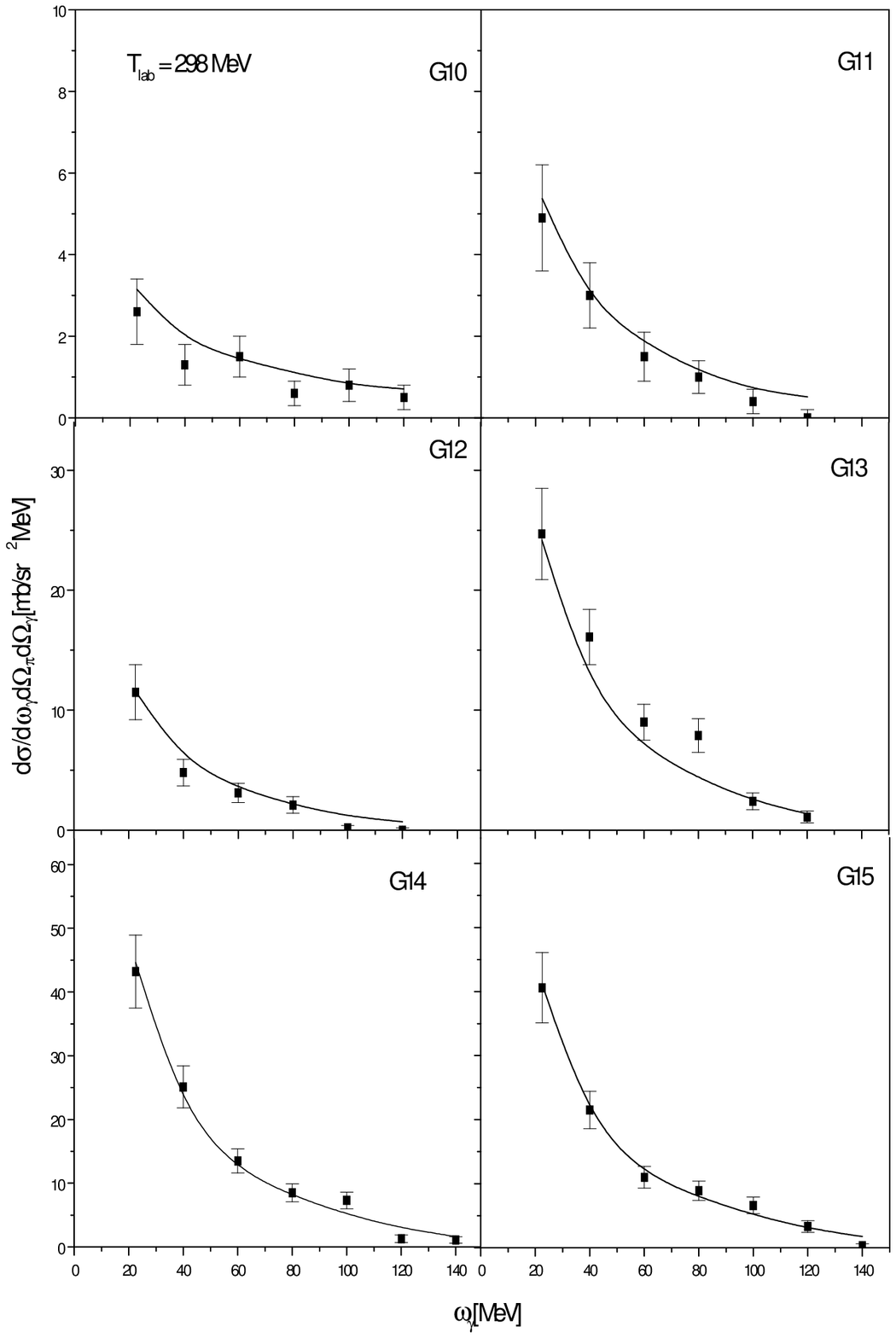}
   \end{center}
\vspace{-3.cm}
\end{figure}
\centerline{\bf\large Figure 11
}
\newpage
\begin{figure}
\begin{center}
    \leavevmode
   \epsfxsize = 15cm
     \epsfysize = 17cm
    \epsffile{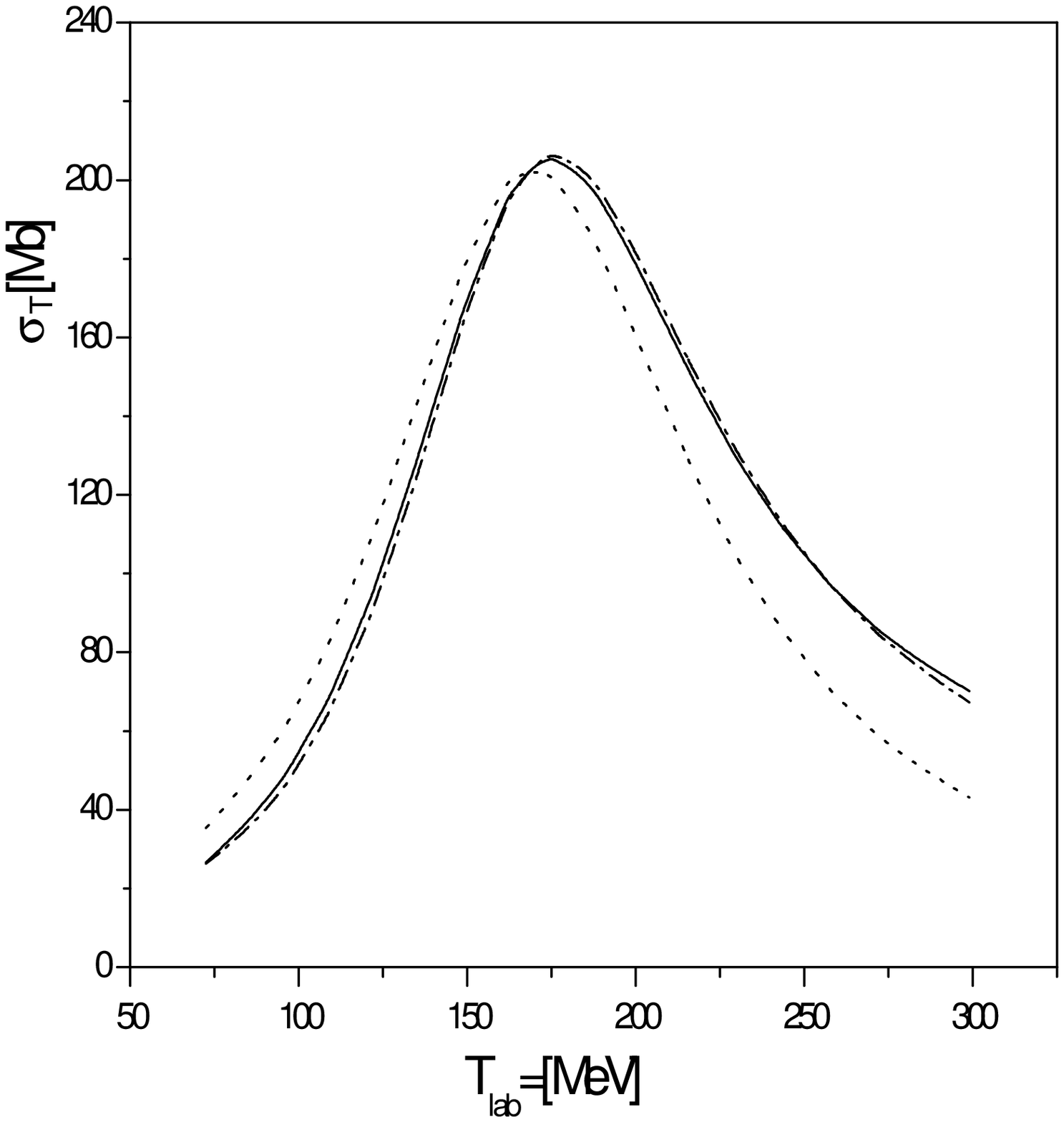}
   \end{center}
\vspace{-3.cm}
\end{figure}
\centerline{\bf\large Figure 12
}


\begin{thebibliography}{99}
\bibitem{pdg98}
D. E. Groom {\it et al}, Particle Data Group, Eur. Phys. J. {\bf C15}, 1
(2000).
\bibitem{Musa}
M. M. Musakhanov, Sov. J. Nuc. Phys. {\bf 19}, 319 (1974).
\bibitem{Hell}
L. Heller, S. Kumano, J.C. Martinez and E. J. Moniz, Phys. Rev.
{\bf C35}, 718 (1987)
\bibitem{ben}
M. Benmerrouche, R.M. Davidson, and Niamai C. Mukhopadhayay, Phys. Rev.
{\bf C39}, 2339 (1989).
\bibitem{wittman}
R. Wittman, Phys. Rev. {\bf C37}, 2075 (1988).
\bibitem{Lin}D. Lin , M.K. Liou and Z.M. Ding, Phys. Rev. {\bf C44},
1819 (1991).
\bibitem{pt}
P. Pascual and R. Tarrach, Nucl. Phys. {\bf B134}, 133 (1978).
\bibitem{sin}
A. Bosshard {\it et al}, Phys. Rev. {\bf D44}, 1962 (1991)
; C. A. Meyer et al., Phys. Rev. {\bf D38}, 754 (1988).
\bibitem{ucla}
B. M. K. Nefkens {\it et al}, Phys. Rev. {\bf D18}, 3911 (1978).
\bibitem{Arman}
M. Arman {\it et al}, Phys. Rev. Lett. {\bf 29}, 962 (1972).
\bibitem{Leung}
K. C. Leung {\it et al}, Phys. Rev. {\bf D14}, 698 (1976).
\bibitem{surdarshan}
K. Johnson and E.C.G. Sudarshan, Ann. Phys. {\bf 13}, 126 (1961)
\bibitem{etemadi}
L. M. Nath, B. Etemadi, and J. D. Kimel, Phys. Rev. {\bf D3}, 2153
(1971); R. E. Behrends and C. Fronsdal, Phys. Rev. {\bf 106}, 277
(1958); J. Ur\'\i as, Ph. D. Thesis, Universit\'e Catholique de
Louvain, Belgium (1976).
\bibitem{Benme}
M. Benmerrouche, R. M. Davidson, and N. C. Mukhopadhyay, Phys. Rev. {\bf
C39}, 2339 (1989); V. Pascalutsa and R. Timmermans, Phys. Rev. {\bf C60},
042201 (1999).
 \bibitem{Schutz}
C. Sch\"utz, J.W. Durso, K. Holinde, and J.
Speth, Phys. Rev. {\bf C49}, 2671(1994).
\bibitem{smatrix}
R. E. Peierls, {\it Proc. of the 1954 Glasgow Conf. on Nucl. and Meson
Physics}, Ed. E. H. Bellamy and R. G. Moorhouse (Pergamon Press, 1955),
p. 296; R. Eden, P. Landshoff, D. Olive, and J. Polkinghorne,
``The Analytic S-matrix" (Cambridge University Press, Cambridge, 1966);
M. L\'evy, Nuovo Cim. {\bf 13}, 115 (1959).
\bibitem{low}
F. E. Low, Phys. Rev. {\bf 110}, 974 (1958).
\bibitem{elamiri}
M. El-Amiri, G. L\'opez Castro and J. Pestieau, Nucl. Phys. {\bf A543},
673 (1992).
\bibitem{nos01}
G. L\'opez Castro and A. Mariano, e-print nucl-th/0006031.
\bibitem{z0}
R. G. Stuart, Phys. Lett. {\bf B262}, 113 (1991); {\bf 272}, 353 (1991);
Phys. Rev. Lett. {\bf 70}, 3193 (1993); A. Sirlin, Phys. Rev. Lett. {\bf
67}, 2127 (1991); H. Veltman, Z. Phys. {\bf C62}, 35 (1994).
\bibitem{fls}
U. Baur and D. Zeppenfeld, Phys. Rev. Lett. {\bf 75}, 1002 (1995); E.
Argyres {\it et
al}, Phys. Lett. {\bf B358}, 339 (1995); M. Beuthe el at Nucl. Phys. {\bf
B498}, 55 (1997); W. Beenakker
{\it et al}, Nucl. Phys. {\bf B500}, 255 (1997).
\bibitem{bls}
G. L\'opez Castro and G. Toledo S\'anchez, Phys. Rev. {\bf D61},
033007 (2000).
\bibitem{Mariano}A. Mariano and G. L\'opez Castro, Phys. Rev. {\bf
62}, 014604(2000).
\bibitem{bernicha}
A. Bernicha, G. L\'opez Castro and J. Pestieau, Nucl. Phys. {\bf A597},
623 (1996).
\bibitem{w}
G. L\'opez Castro, J. L. Lucio  M.,  and J. Pestieau, Mod. Phys. Lett.
{\bf A}, (1991); Int. J. Mod. Phys. {\bf A10}, (1996).
\bibitem{pn93}
A. Pilaftsis and M. Nowakowski, Z. Phys. {\bf C60}, 121 (1993).
\bibitem{sigma}See for example: B.C. Pearce and B.K. Jennings, Nucl.
Phys.,{\bf A528},655 (1991); C. Lee, S.N. Yang, and T.-S.H. Lee, J. Phys.
{\bf G17}, L131 (1991).
\bibitem{pedroni} E. Pedroni {\it et al}, Nucl. Phys. {\bf A300},
321(1978).
\bibitem{Bussey} P.J. Bussey {\it et al}, Nuc. Phys. {\bf B58},
363(1973).
\bibitem{Sadler} M.E. Sadler {\it et al}, Phys. Rev. {\bf
D35}, 2718 (1987).
\bibitem{Gordeev} V.A. Gordeev {\it et al}, Nuc. Phys. {\bf A364},
408 (1981).
\bibitem{bk} T. H. Burnett and N. M. Kroll, Phys. Rev. Lett. {\bf 20}, 86
(1968).
\bibitem{kpz}
V. I. Zakharov, L. A. Kondratyuk, and L. A. Ponomarev, Yad. Fiz. {\bf
8}, 783 (1968) [Sov. J. Nucl. Phys. {\bf 8}, 456 (1968)].
\bibitem{blp}
M. A. B. Beg, B. W. Lee and A. Pais, Phys. Rev. Lett. {\bf 13}, 514
(1964).
\bibitem{krivo}
M. I. Krivoruchenko, Sov. Jour. Nucl. Phys. {\bf 45}, 109 (1987).
\bibitem{bp}
M. A. B. Beg and A. Pais, Phys. Rev. {\bf 137}, B1514 (1965)
\bibitem{brv}
G. E. Brown, M. Rho and V. Vento, Phys. Lett. {\bf B97}, 423 (1980).
\bibitem{Fra}
J. Franklin, e-print hep-ph/0103139.
\end{thebibliography}
\end{document}